\documentclass[aps,prx,twocolumn,superscriptaddress]{revtex4-2}

\usepackage{graphicx}
\usepackage[colorlinks]{hyperref}
\hypersetup{
    colorlinks=true,
    linkcolor=blue,
    filecolor=blue,      
    urlcolor=blue,
    citecolor=blue
}

\usepackage{enumitem}
\setlist[enumerate]{leftmargin=*}

\usepackage{physics}
\usepackage{siunitx}

\usepackage{amssymb}
\usepackage{mathtools}
\newcommand{\myvar}[1]{\sigma_{#1}^2}
\newcommand{\conj}[1]{#1^*}
\newcommand{\numberthis}{\addtocounter{equation}{1}\tag{\theequation}}
\newcommand{\hc}{\text{H.c.}}
\newcommand{\iu}{\text{i}}
\newcommand{\e}{e}

\newcommand{\hilbert}{\mathcal{H}}
\DeclareMathOperator{\rect}{rect}
\DeclareMathOperator{\sign}{sign}
\DeclareMathOperator{\sinc}{sinc}
\DeclareMathOperator{\arccosh}{arccosh}

\DeclareMathOperator{\erfc}{erfc}
\DeclareMathOperator{\erfcx}{erfcx}
\DeclareMathOperator{\cov}{cov}

\begin{document}

\title{Optical time-domain quantum state tomography on a subcycle scale}
\author{Emanuel Hubenschmid}
\email[]{emanuel.hubenschmid@uni-konstanz.de}
\affiliation{Department of Physics, University of Konstanz, D-78457 Konstanz, Germany}

\author{Thiago L. M. Guedes}
\email[]{thiago.lucena@uni-konstanz.de}
\affiliation{Department of Physics, University of Konstanz, D-78457 Konstanz, Germany}
\affiliation{Institute for Quantum Information, RWTH Aachen University, D-52056 Aachen, Germany}
\affiliation{Peter Gr{\"u}nberg Institute, Theoretical Nanoelectronics, Forschungszentrum J{\"u}lich, D-52425 J{\"u}lich, Germany}

\author{Guido Burkard}
\email[]{guido.burkard@uni-konstanz.de}
\affiliation{Department of Physics, University of Konstanz, D-78457 Konstanz, Germany}

\begin{abstract}
	Following recent progress in the experimental application of electro-optic sampling to the detection of the quantum fluctuations of the electromagnetic-field ground state and ultrabroadband squeezed states on a subcycle scale, we propose an approach to elevate broadband electro-optic sampling from a spectroscopic method to a full quantum tomography scheme, able to reconstruct a broadband quantum state directly in the time-domain. By combining two recently developed methods to theoretically describe quantum electro-optic sampling, we analytically relate the photon-count probability distribution of the electro-optic signal to a transformed phase-space quasiprobability distribution of the sampled quantum state as a function of the time delay between the sampled mid-infrared pulsed state and an ultrabroadband near-infrared pump/probe pulse. We catalog and analyze sources of noise and show that in quantum electro-optic sampling with an ultrabroadband pump pulse one can expect to observe thermalization due to entanglement breaking. Mitigation of the thermalization noise enables a tomographic reconstruction of broadband quantum states while granting access to its dynamics on a subcycle scale.
\end{abstract}

\maketitle

\section{Introduction}
\begin{figure}[t!]
    \includegraphics[width=0.48\textwidth]{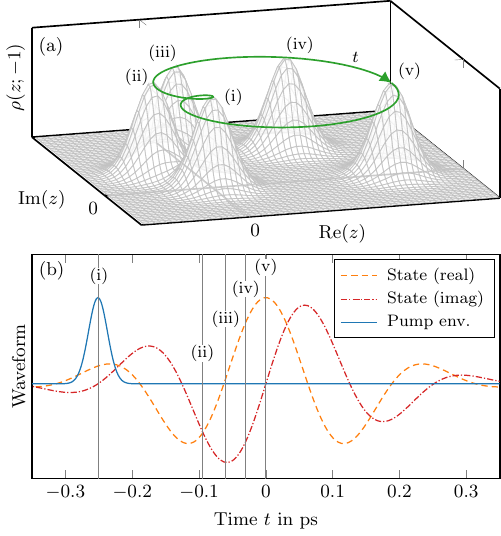}
	\caption{The goal of time-domain quantum tomography is to reconstruct the quantum state and dynamics of a pulsed quantum state directly in the time domain. (a) Exemplary time-evolution in phase space of the broadband (Husimi) quasiprobability distribution $\rho(z;-1)$ of a coherent (pulsed) quantum state, whose quadratures evolve dynamically following the solid green line. The free time evolution of the distribution starts in the vacuum at (i), spirals outwards via (ii)-(iv) until reaching maximum displacement in (v) before returning to the vacuum. (b) The corresponding sampled waveform of the pulsed quantum state with the dashed orange (dash-dotted red) line for the real (imaginary) part. The state is sampled and thus averaged over the duration of a pump pulses envelope (solid blue). If the pump pulse is shorter than the duration of a single cycle of the sample pulse, the measurement is said to be subcycle. For a pump pulse centered at the vertical line located at (i), the sampled waveform averaged over the duration of the pump is close to zero, while at (v) the average real part of the waveform is maximal, matching the free time evolution in (a).}
	\label{fig:cartoon}
\end{figure}

Accessing the dynamics of broadband time-dependent quantum states remains challenging even for state-of-the-art experiments. While many theoretical descriptions of quantum optical systems resort to monochromatic modes, in practice, non-monochromatic modes are unavoidable for many applications~\cite{gulla2021}. Wave packets, for example, are inherently multi-mode and are vital for transmitting quantum information over long distances and thus necessary for quantum communication technologies \cite{Weedbrook2012,Braunstein2005} such as continuous-variable quantum key distribution \cite{Weedbrook2004,Lance2005,Madsen2012,Usenko2015,Diamanti2015,Hosseinidehaj2019}. For these technologies it is crucial to prepare the correct target state, which is usually validated using homodyne quantum tomography \cite{Vogel1989,Smithey1993,Leonhardt1994,Wallentowitz1996,Breitenbach1997,Luis2015,Bohmann2018,Tiedau2018,Knyazev2018,Olivares2019,Walker1986,Freyberger1993,Leonhardt1993,Zucchetti1996,Rehacek2015}. To obtain not only the amplitude, but also the phase information of the sampled optical state, homodyne tomography relies on the interference with a local oscillator whose temporal modes overlap partially or entirely with those of the sampled state within a beam splitter \cite{Slusher1987,Hirano1990,Smithey1992,Smithey1993,Smithey1993a,Zavatta2002,Zavatta2005,Haderka2009,Okubo2008}. The large overlap of the sampled pulsed mode with the local oscillator unavoidably leads to an averaging of the measured signal over the detection time. In addition, homodyning generally leads to the destruction of the sampled state. The averaging effect can be reduced by the use of a ultrashort local-oscillator pulse, such that the sampled state is only averaged over the envelope of the short pulse, as shown schematically in Fig.~\ref{fig:cartoon}. Furthermore, the destruction of the sampled state can be avoided by indirect detection. These modifications are defining characteristics of electro-optic sampling \cite{Namba1961,Gallot1999,Leitenstorfer1999,Sulzer2020,Moskalenko2015,Guedes2019,Kizmann2019,BeneaChelmus2019,Lindel2020,Lindel2021,Onoe2022,Kizmann2022,Guedes2023,Settembrini2022,Beckh2021}.
Electro-optic sampling utilizes a short, higher-frequency ancillary pulse, usually in the near-infrared (NIR), to sample a longer pulse of lower frequency, usually in the mid-infrared (MIR) range. The ellipticity of the higher-frequency modes is correlated with the state of the lower-frequency modes, averaged over a small time slice defined by the ancillary pulse. Therefore, by measuring the change in ellipticity of the high-frequency modes one can infer the state of the low-frequency modes at the time of interaction. Such measurements require the usage of ancillary pulses shorter than the duration of a single cycle of the sampled broadband state, the so-called subcycle regime.

There are two approaches to optical quantum tomography. One is based on the simultaneous measurement of two noncommuting electromagnetic-field quadratures, which allow to reconstruct the phase-space quasiprobability distribution of the quantum state, and therewith the quantum state, directly from the sampled data \cite{Walker1986,Freyberger1993,Leonhardt1993,Zucchetti1996,Rehacek2015}. In the other approach, the quantum state is reconstructed from single-quadrature measurements using the Radon-transform \cite{Vogel1989,Smithey1993,Breitenbach1997}. These state-of-the-art quantum tomography methods to reconstruct the waveform and the quantum state of the electromagnetic field do not operate in the subcycle regime and thus require the sampled state to be composed of at most a few reproducible orthogonal mode profiles \cite{Tiedau2018,Ansari2017,Ansari2018,GilLopez2021}. Recently, electro-optic sampling has been successfully used to sample the statistics of the electromagnetic-field broadband ground state \cite{Riek2015}, as well as of ultrabroadband squeezed states in the time domain with subcycle resolution \cite{Riek2017}. On the theory side, new tools have been developed to account for quantum contributions to the nonlinear interaction necessary to generate the electro-optic signal \cite{Moskalenko2015,Guedes2019,Kizmann2019,BeneaChelmus2019,Scheel1998,Knoll2003,Scheel2009,Virally2019,Virally2021,Lindel2020,Lindel2021,Onoe2022,Kizmann2022,Guedes2023,Settembrini2022}. However, a description of the full photon-counting statistics of the electro-optic signal for the measurement of broadband quantum states and its relation to phase-space distributions are still missing. Here, we close this conceptual gap by merging two recently developed theoretical methods \cite{Onoe2022,Hubenschmid2022} to derive the probability distribution of the quantum-electro-optic measurement outcomes. We propose a modified quantum-electro-optic sampling setup capable of measuring two noncommuting quadratures of the spatiotemporally localized MIR mode simultaneously: Instead of one detection stage, two detection stages sensitive to two neighboring non-overlapping spectral cuts of the high-frequency signal generated by the nonlinear interaction are used. We utilize the method of first-order unitary developed in Ref.~\cite{Onoe2022} and simplify the description to involve solely a few effective nonmonochromatic modes. We can then apply the framework developed in Ref.~\cite{Hubenschmid2022} and relate the count-probability distribution of the electro-optic signal to a (transformed) phase-space quasiprobability distribution. The transformation is due to entanglement breakage during the sampling process and leads to excess noise. We propose a means to attenuate this noise by selecting a frequency band reducing the entanglement breakage prior to detection. The resulting model allows for the description of multichannel quantum electro-optic sampling, and we show how this proposed measurement scheme can be used for optical time-domain quantum tomography with subcycle resolution.

Our paper is structured as follows.
In Sec.~\ref{sec:theoretical_description} we develop a theoretical model to describe quantum electro-optic sampling. Sec.~\ref{sec:result} presents the main result of this work: The count-probability distribution of the electro-optic signal. In Sec.~\ref{sec:tomography} we utilize the analysis developed in the preceding sections to discuss realistic implementations of optical time-domain quantum tomography.

\section{Theoretical framework}\label{sec:theoretical_description}
Before embarking on the formal theoretical description of our model, we develop an intuitive picture of electro-optic sampling as a quantum tomography scheme for reconstructing the quantum state of the electromagnetic field with a subcycle resolution in Sec.~\ref{sec:heuristics}. In the subsequent sections, we then describe the formal model of our proposed generalization of quantum electro-optic sampling. We divide the measurement protocol into three stages: the nonlinear interaction (Sec.~\ref{sec:NL}), the filtering stage (Sec.~\ref{sec:filtering}) and the ellipsometry stage (Sec.~\ref{sec:ellip}). In Sec.~\ref{sec:first_order} we show how our model improves and generalizes the effective description of the nonlinear interaction laid down in Ref.~\cite{Onoe2022}.

\subsection{Heuristics of (time-domain) quantum tomography}\label{sec:heuristics}
In optical tomography of continuous-variable quantum states the reconstruction of the state is usually performed in phase space. A (single-mode) quantum state can be represented by a distribution over a complex argument, where the real part corresponds to the generalized coordinate (X quadrature) and the imaginary part to the generalized momentum (Y quadrature) of the mode. One example of such a quasiprobability distribution is the widely known Wigner function; another related example is the Husimi distribution, visualized for a broadband coherent state in Fig.~\ref{fig:cartoon} (a) at different points of its time evolution. With a finite number of measurements, these distributions can only be sampled at discrete points, which for homodyne detection and electro-optic sampling are in practice determined by the difference counts $\Delta n_i$ of two balanced photon detectors, $i=\text{X},\text{Y}$. Over time the quasiprobability distribution of a single-mode quantum state would rotate in phase space and thus by directly measuring the photon number in that single mode, only information about the amplitude, but not about the phase, can be collected. For this reason, an ancillary mode in a coherent state $\beta$ is introduced, also referred to as a local oscillator, with which the single mode can interfere. If these balanced detections make use of local oscillators shifted by $\pi/2$ relative to each other, the tomography scheme is able to sample the quasiprobability distribution of the single mode at the complex arguments $z(\{\Delta n_i\}) = (\Delta n_\text{X} + \iu \Delta n_\text{Y})/\abs{\beta}$ \cite{Walker1986,Freyberger1993,Leonhardt1993,Zucchetti1996,Rehacek2015}. Although popularly used tomography techniques like homodyne detection make use of direct measurements, entangling the analyzed mode to ancillary modes using a nonlinear crystal implements an indirect measurement, as we have shown in \cite{Hubenschmid2022}. In this case, the difference-count probability distribution
\begin{equation}\label{eq:heuristics}
	p(\{\Delta n_i\}) \propto \rho\left[z(\{\Delta n_i\})/\sinh(\abs{\zeta}); \tilde{s}\right]
.\end{equation}
is proportional to the $\tilde{s}$-quasiprobability distribution $\rho$ of the quantum state in which the parameter $\tilde{s} = 1 - 2\coth^2(\abs{\zeta})$ is determined by the two-mode squeezing strength $\zeta$ of the nonlinear crystal. These distributions are a smoothed-out version of the Wigner function, which is achieved by convolving the Wigner function with a Gaussian distribution of variance proportional to $-\tilde{s}$. To ensure non-negativity of the quasiporbability distribution, one requires $\tilde{s}<-1$. With increasing squeezing strength, the measurement is made stronger and $\tilde{s}$ approaches its maximum value of $-1$.

If the nonlinear crystal is driven by an ultrabroadband pulse instead of a few monochromatic modes, the tomography scheme from the previous paragraph constitutes a (generalized) version of quantum electro-optic sampling. Electro-optic sampling utilizes the electro-optic effect to sample a short low-frequency pulse using an even shorter high-frequency pulse for which the sampled pulse seems quasi-instantaneous. Because of the electro-optic (Pockels) effect, the electric field of the sampled low-frequency modes induces (or changes) birefringence in the nonlinear crystal which leads to a change in the ellipticity of the probing high-frequency pulse at the time of interaction \cite{Boyd2019}. The ellipticity can then be measured using a balanced photon-detection between the two polarization directions of the high-frequency pulse (ellipsometer) and thus the state of the low-frequency electric field averaged over the time of interaction can be inferred \cite{Namba1961,Gallot1999,Leitenstorfer1999,Sulzer2020}. By repeating this measurement for different time delays $\Delta t$ between the high- and low-frequency pulse, the electric field can be sampled as a function of time (delay), as depicted in Fig.~\ref{fig:cartoon} (b). In order to sample the waveform directly in the time domain, the high-frequency pulse needs to be shorter than a single optical cycle of the low-frequency pulse. If this is the case, the measurement is said to be subcycle. In principle the same could be achieved with homodyne detection, although this would require either fast photodetectors or the local oscillator to be highly subcycle with respect to itself. The latter requirement is due to the shared frequency domain of the local oscillator and the sampled pulse in homodyne detection. The nonlinear mechanism underlying electro-optic sampling allows the high-frequency pulse to be single-cycle or even a few cycles long, while retaining its subcycle character relative to the low-frequency pulse. Furthermore, electro-optic sampling makes accessible the frequency range right between the gap of electronic and optical frequencies, the so-called mid-infrared range. Generation and detection of mid-infrared frequencies using electronics or optics remain challenging. On the other hand, detection in the high-frequency range, usually spanning from near infrared to optical frequencies, can be easily achieved using optical components.

If the probed low-frequency modes are in a quantum state, we would expect the count-probability distribution to follow the quasiprobability distribution of the quantum state averaged over the time window defined by the short probing high-frequency pulse, as sketched in Fig.~\ref{fig:cartoon} (b). However, quantum mechanically the short high-frequency pulse has an additional quantum mechanical effect. Since the Gabor limit forces a bandwidth broadening if the pulse is made shorter, additional unsampled modes get entangled to the sampled ones. Therefore, the entanglement between the sampled and unsampled modes is broken through measurement. This entanglement breakage leads to an increase of the von Neumann entropy of the sampled quantum state, which in turn creates thermal excitations \cite{Onoe2022}. The thermalization is reflected in the electro-optic signal as excess noise [see Fig.~\ref{fig:sfg_effect} and Sec.~\ref{sec:result}], leading to a resolution tradeoff between time and phase-space coordinates. The sampled frequencies are up-converted to the detected modes via difference-frequency generation, while the thermalized modes are up-converted by sum-frequency generation. Therefore, the excess noise due to thermalization can be mitigated by filtering the high-frequencies below the central frequency of the high-frequency pump pulse.

\subsection{The nonlinear interaction}\label{sec:NL}
Since electro-optic sampling is an indirect measurement of low-frequency modes mediated by a subcycle pulse of high-frequency ancillary modes, the two frequency ranges have to be correlated in order to access information about the former by measuring the latter. In this work we consider low-frequency mid-infrared (MIR) and high-frequency near-infrared (NIR) modes, which interact in a zinc-blende-type nonlinear crystal of length $L$, refractive index $n_{\omega}$ [given by Eq.~\eqref{eq:refractive_index}] and a coupling constant $d = -n_{\omega_{\text{p}}}^4 r_{41}$ dependent on the refractive index at the central pump frequency $\omega_{\text{p}}$ and the second-order electro-optic (susceptibility) coefficient, $r_{41} = \SI{4}{\pico\meter\per\volt}$ (see \cite{Boyd2019}, p.~500).
\begin{figure}
	\includegraphics[width=0.45\textwidth]{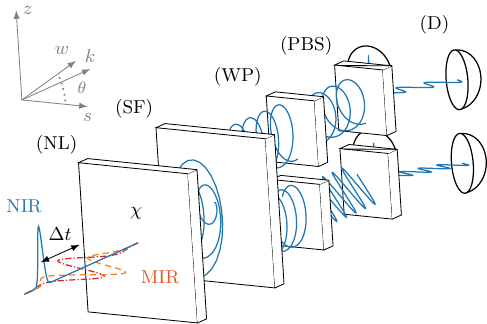}
	\caption{Schematic of the proposed electro-optic quantum tomography setup. The $z$-polarized near-infrared (NIR) pulse interacts with the $s$-polarized mid-infrared (MIR) temporal mode in the zinc-blende-type nonlinear crystal, labeled (NL), and generates new excitations in the $s$-polarized NIR modes. By varying the time delay $\Delta t$ between the NIR pulse and the pulsed MIR mode, the MIR mode is sampled. A polarizing beam-splitter can be used to remove the $z$-polarized pump and a new probe pulse with a different amplitude can be introduced as a replacement. The $s$- and $z$-polarized NIR modes then undergo a spectral filter station, labeled (SF). Two narrow (quasimonochromatic) and non-overlapping frequency bands $\tilde{\omega}$ of the NIR modes are selected, each being directed towards one detection stage. The two NIR quasimonochromatic modes with frequency bands $\Delta \omega_i$ pass through a $\varphi_i$ waveplate rotated by an angle $\theta_i$, labeled (WP). Afterwards the $s$ and $z$ components of each quasimonochromatic NIR frequency band are split spatially with the aid of a polarizing beam splitter, labeled (PBS). The photon numbers $\hat{n}_{i,s}$ and $\hat{n}_{i,z}$ in the $s$ and $z$ polarized components of each quasimonochromatic cut are counted using photo detectors, labeled (D), and the difference $\Delta \hat{n}_i = \hat{n}_{i,s} - \hat{n}_{i,z}$ constitutes the electro-optic signal.} 
	\label{fig:setup}
\end{figure}
As depicted in Fig.~\ref{fig:setup}, in the non-linear crystal (NL), the interaction of the $z$-polarized NIR modes with the $s$-polarized MIR modes generates new excitations in the $s$-polarized NIR modes (for details about the geometric arrangement see Ref~\cite{Moskalenko2015}). This interaction can be described by the unitary time-evolution operator $\hat{U}_\text{NL} = \exp(\hat{S}_\text{NL})$ with \cite{Guedes2019,Onoe2022}
\begin{equation}\label{eq:NL_action}
	\hat{S}_\text{NL} = \int_{\abs{\Omega}<\Lambda} \int_{\Lambda < \abs{\omega}} S(\Omega,\omega) \hat{a}_{\Omega,s} \hat{a}_{\omega,s}^\dagger \dd \Omega \dd \omega
,\end{equation}
as long as the coherent pump is strong compared to any quantum contributions (including depletion), absorption is negligible (off-resonant regime) and there is no overlap between the MIR and NIR frequency ranges of interest. To ensure skew Hermiticity of $\hat{S}_\text{NL}$, the condition $\conj{S}(\omega,\Omega) = -S(\Omega,\omega)$ has to be fulfilled. We denote angular frequencies in the MIR range by $\Omega$ while the NIR range is represented by $\omega$. From here on we will refer to the angular frequency just as frequency. To avoid frequency crossing between the NIR and MIR ranges, a cutoff frequency $\Lambda$ is introduced. The operator $\hat{a}_{\omega,s}$ with $\omega>0$ is the annihilation operator for the mode with frequency $\omega$ and polarization $s$. We use the convention $\hat{a}_{-\omega,s} = \hat{a}^\dagger_{\omega,s}$ and $\comm*{\hat{a}_{\omega,s}}{\hat{a}^\dagger_{\omega^\prime,s}} = \sign(\omega)\delta(\omega - \omega^\prime)$. The joint spectral amplitude
\begin{equation}\label{eq:two_mode_spectrum}
	S(\Omega,\omega) = [\alpha_\text{p} E_\text{p}(\Omega-\omega) + \conj{\alpha}_\text{p} \conj{E}_\text{p}(\omega-\Omega)]\zeta_{\Omega,\omega}
,\end{equation}
is determined by two components. First, by the phase-matching function,
\begin{align}
	&\zeta_{\Omega,\omega} = -\iu d \sign(\omega\Omega)\sqrt{\frac{\abs{\omega\Omega}}{n_{\Omega}n_{\omega}}}\frac{L}{2c}\sinc(\eta_{\omega,\Omega}), \label{eq:phase_matching} \\
	&\eta_{\Omega,\omega} = \frac{L}{2c}[\omega(n_{\omega} - n_{\omega-\Omega}) - \Omega(n_{\Omega} - n_{\Omega-\omega})] \label{eq:eta}
,\end{align}
including the speed of light $c$. Second, by the coherent $z$-polarized ultrabroadband NIR pump of amplitude $\alpha_\text{p}$, beam-waist area $A$, and spectrum
\begin{equation}\label{eq:pump_field}
	E_\text{p}(\omega) = \iu\left(\frac{\hbar}{4\pi c \varepsilon_0 A}\right)^{\frac{1}{2}}\sqrt{\frac{\abs{\omega}}{n_\omega}} f_\text{p}(\omega)
,\end{equation}
with the vacuum permittivity $\varepsilon_0$ and the reduced Planck constant $\hbar$. The mode function $f_\text{p}(\omega)$ of the pump is assumed to be a Gaussian
\begin{align}
	&f_\text{p}(\omega) = N_\text{p} \exp[-(\omega - \omega_\text{p})^2/(2\sigma_\text{p})^2 - \iu t_\text{p}\omega], \label{eq:probe_spec} \\
	&N_\text{p} = \left\{\sqrt{\frac{\pi}{2}}\sigma_\text{p} \left[\erfc\left(\frac{-\omega_\text{p}}{\sqrt{2}\sigma_\text{p}}\right) - \erfc\left(\frac{\omega_\text{p}}{\sqrt{2}\sigma_\text{p}}\right)\right]\right\}^{-\frac{1}{2}}
\end{align}
of bandwidth $\sigma_\text{p}$ and central frequency $\omega_\text{p}$. The normalization constant $N_\text{p}$ ensures that $\int_{-\infty}^\infty \sign(\omega)\abs{f_\text{p}(\omega)}^2 \dd\omega = 1$. 

A closer inspection of Eq.~\eqref{eq:two_mode_spectrum} lets us differentiate four frequency domains. If both the high and the low frequencies are positive ($\omega > 0$ and $\Omega > 0$) the first summand in the right-hand side of Eq.~\eqref{eq:two_mode_spectrum} is exponentially suppressed due to $E_\text{p}(\Omega -\omega)$ selecting $\Omega$ frequencies far beyond the MIR. Consequently, only $\conj{E}_\text{p}(\omega -\Omega)$ contributes, favoring frequencies obeying $\abs{\omega} \sim \omega_{\rm p} + \abs{\Omega}$. The detected $s$-polarized high-frequency excitations therefore correspond to sum-frequency generation (SFG). The conjugate process ($\omega < 0$ and $\Omega < 0$), on the other hand, suppresses the second summand in Eq.~\eqref{eq:two_mode_spectrum}. SFG together with its conjugate process result in a beam-splitter-like contribution to Eq.~\eqref{eq:NL_action}. If the high and low frequency have opposite signs (i.e., $\omega > 0$ and $\Omega < 0$ or $\omega < 0$ and $\Omega > 0$), the detected high-frequencies can be associated with difference-frequency generation (DFG) and modes fulfilling $\abs{\omega} \sim \omega_{\rm p} - \abs{\Omega}$ are favored. Again, one of the summands in the right-hand side of Eq.~\eqref{eq:two_mode_spectrum} is exponentially suppressed. Considering DFG together with its conjugate process results in a squeezing-like contribution to Eq.~\eqref{eq:NL_action}. For a detailed derivation of the nonlinear unitary operator, Eq.~\eqref{eq:NL_action}, see Ref.~\cite{Onoe2022}.

\subsection{The filtering stage}\label{sec:filtering}
To allow for a more versatile description, we assume that pump and probe pulses can differ: After the NIR pulse propagates through the nonlinear crystal, a polarizing beam-splitter can be used to remove the $z$-polarized pump and a new coherent probe pulse with amplitude $\beta$ can be introduced as a replacement. The combined $s$- and $z$-polarized NIR modes are then spectrally filtered (e.g., by a band-pass filter) to give a narrow quasimonochromatic profile centered at $\tilde{\omega}$ and of bandwidth $\Delta \omega$. The filtering gives some control over the ratio between DFG and SFG contribution to the signal. If frequencies above the central frequency of the pump are detected, $\tilde{\omega} > \omega_\text{p}$, mainly SFG contributes to the signal while for the opposite case, $\tilde{\omega} < \omega_\text{p}$, DFG is dominant \cite{Sulzer2020}. The filtered probe together with the filtered $s$-polarized NIR excitations generated by the nonlinear interaction are split spectrally into multiple narrow-band pulses, as indicated in Fig.~\ref{fig:setup} (SF). This allows one to measure different quadratures of the MIR modes simultaneously using the different quasimonochomatic spectral splits. The new pulses $i \in I = \{\text{X},\text{Y}\}$ are characterized by a central frequency $\tilde{\omega}_i$ and a bandwidth $\Delta \tilde{\omega}_i$ and define a set of discrete modes $\hat{\mathfrak{u}}_{\tilde{\omega}_i,\lambda} = (\Delta \omega_i)^{-\frac{1}{2}} \int_{-\infty}^\infty \rect\left(\frac{\tilde{\omega}_i-\omega}{\Delta \omega_i}\right)\hat{a}_{\omega,\lambda} \dd \omega$. The rectangular function $\rect(x)$ is equal to $1$ for $\abs{x} < 1/2$,  equal to $1/2$ for $\abs{x} = 1/2$, and $0$ for $\abs{x} > 1/2$. To express the total NIR mode operator
\begin{equation}\label{eq:spectral_splitting}
	\hat{\mathfrak{u}}_{\tilde{\omega},\lambda} = \int_{-\infty}^\infty \rect\left(\frac{\tilde{\omega}-\omega}{\Delta \omega}\right) \hat{a}_{\omega,\lambda} \dd \omega = \sum_{i \in I} \tilde{\alpha}_i \hat{\mathfrak{u}}_{\tilde{\omega}_i,\lambda}
\end{equation}
in terms of the set of discrete mode operators, we have to choose $\tilde{\alpha}_i = \sqrt{\Delta \omega_i/\Delta \omega}$ and require $\sum_{i \in I} \rect\left(\frac{\tilde{\omega}_i-\omega}{\Delta \omega_i}\right) = \rect\left(\frac{\tilde{\omega}-\omega}{\Delta \omega}\right)$. The latter condition ensures that both spectral cuts are correlated through the nonlinear interaction with practically the same MIR frequency range (and therefore the same nonmonochromatic mode). We will denote discretized mode operators, defined over a frequency range, with $\mathfrak{u}$, $\mathfrak{a}$, etc. This distinction is important because the usual discrete-mode commutation relations are violated by pairs of such operators defined over overlapping frequency bands, and we therefore assume throughout this work that detected frequency bands are so selected that no such overlaps can happen. The replacement of the pump by a probe pulse together with the frequency splitting of the $z$-polarized modes can be described by the displacement operators $\hat{D}_{\tilde{\omega}_i,z}(\beta_i) = \exp(\beta_i \hat{\mathfrak{u}}_{\tilde{\omega}_i,z}^\dagger - \hc)$ with amplitudes $\beta_i = \tilde{\alpha}_i\beta$, where $\beta$ is the post-filtered probe amplitude.

Alternatively, the simultaneous measurement can be realized using a beam splitter instead of the spectral splitting, resulting in the same decomposition in the last equality of Eq.~\eqref{eq:spectral_splitting}, as shown in Appendix \ref{a:two_mode_char}. This alternative approach upgrades the electro-optic-sampling setup proposed in Ref.~\cite{Sulzer2020} to a quantum-tomography protocol.

\subsection{The ellipsometry stage}\label{sec:ellip}
Subsequent to the filtering stage, we assume that an ellipsometry on each NIR mode $\hat{\mathfrak{u}}_{\tilde{\omega}_i,\lambda}$ is performed: A $\phi_i$-wave plate rotated by an angle $\theta_i$, (WP) in Fig.~\ref{fig:setup}, makes the signal at each ellipsometry stage balanced. We choose the signal to be balanced, i.e., to yield zero average signal for an MIR state for which all quadrature expectation values vanish, because noise affecting both polarizations the same way cancels out (see \cite{Hubenschmid2022}, Sec.~III for details). The action of the wave plate is assumed to be independent of the frequency on each frequency band $\Delta \omega_i$ and can thus be modeled by $\hat{U}_{\tilde{\omega}_i,\text{WP}} = \exp(\iu \phi_i \hat{\mathfrak{u}}_{\tilde{\omega}_i,\theta_i}^\dagger \hat{\mathfrak{u}}_{\tilde{\omega}_i,\theta_i})$, where $\hat{\mathfrak{u}}_{\tilde{\omega}_i,\theta_i} = \cos(\theta_i)\hat{\mathfrak{u}}_{\tilde{\omega}_i,s} + \sin(\theta_i)\hat{\mathfrak{u}}_{\tilde{\omega}_i,z}$ is the bosonic operator acting on the mode with polarization parallel to the optical axis of the wave plate. The total time evolution of the probe-and-sample system is described by
\begin{equation}\label{eq:total_time_evolution}
	\hat{U} = \hat{U}_{\text{WP}} \hat{D}_{\tilde{\omega}}(\vec{\beta})\hat{U}_\text{NL}
,\end{equation}
with $\hat{U}_{\text{WP}} = \bigotimes_i \hat{U}_{\tilde{\omega}_i,\text{WP}}$ and $\hat{D}_{\tilde{\omega}}(\vec{\beta}) = \bigotimes_i \hat{D}_{\tilde{\omega}_i,z}(\beta_i)$. After the waveplate, the $s$- and $z$-polarized photons are spatially separated with the aid of polarizing beam splitters, (PBS) in Fig.~\ref{fig:setup}. The photon number in each polarization of every frequency cut $\tilde{\omega}_i$ is measured using photon detectors, (D) in Fig.~\ref{fig:setup}. The electro-optic signals are the photon-count differences described by
\begin{equation}\label{eq:count_diff_cont}
	\Delta \hat{n}_i = \int_{-\infty}^\infty \rect\left(\frac{\tilde{\omega}_i-\omega}{\Delta \omega_i}\right)\left(\hat{n}_{\omega,s} - \hat{n}_{\omega_i,z}\right) \dd \omega
.\end{equation}
By exploiting the narrow bandwidth of the operators $\hat{\mathfrak{u}}_{\tilde{\omega}_i,\lambda}$, we can approximate the photon number,
\begin{equation}\label{eq:narrow_photon_number}
	\hat{\mathfrak{n}}_{\tilde{\omega}_i,\lambda} = \int_{-\infty}^\infty \rect\left(\frac{\tilde{\omega}_i-\omega}{\Delta \omega}\right) \hat{a}_{\omega,\lambda}^\dagger\hat{a}_{\omega,\lambda} \dd \omega \approx \hat{\mathfrak{u}}_{\tilde{\omega}_i,\lambda}^\dagger\hat{\mathfrak{u}}_{\tilde{\omega}_i,\lambda}
,\end{equation}
and thus express the photon-number difference in terms of the photon number in each (polarization) mode with frequency $\tilde{\omega}_i$,
\begin{equation}\label{eq:count_diff_approx}
	\Delta \hat{n}_i \approx \hat{\mathfrak{n}}_{\tilde{\omega}_i,s} - \hat{\mathfrak{n}}_{\tilde{\omega}_i,z} = \sum_{\Delta n_i = -\infty}^\infty \Delta n_i \hat{P}_{\Delta n_i}
.\end{equation}
In the last step of the above equation, we decomposed the observable $\Delta \hat{n}_i$ into projectors
\begin{equation}\label{eq:projectors}
	\hat{P}_{\Delta n_i} = \sum_{n_i=\tilde{n}_i}^\infty\ket{n_i+\Delta n_i}_{i,s}\prescript{}{i,s}{\bra{n_i+\Delta n_i}} \otimes \ket{n_i}_{i,z}\prescript{}{i,z}{\bra{n_i}}
,\end{equation}
onto the eigenspace of NIR Fock states with a photon-number difference $\Delta n_i$ between the $s$- and $z$-polarized component. The summation starts at $\tilde{n}_i=\max\{0,-\Delta n_i\}$ to avoid negative photon-number counting.

The projection operator in Eq.~\eqref{eq:projectors}, as well as the total time-evolution operator in Eq.~\eqref{eq:total_time_evolution}, with the exception of the nonlinear unitary $\hat{U}_{\rm NL}$, are equivalent to the corresponding operators used in Ref.~\cite{Hubenschmid2022}. While the unitary defined by Eq.~\eqref{eq:NL_action} describes a squeezing interaction between a continuum of modes, the unitary operator
\begin{equation}\label{eq:two_mode_squeezing}
	\hat{U}_\text{NL,MONO} = \exp(\conj{\zeta}\hat{a}_{\Omega,s}\sum_{i \in I} \tilde{\alpha}_i \hat{a}_{i,s}-\hc)
,\end{equation}
used in Ref.~\cite{Hubenschmid2022}, describes a squeezing interaction between a monochromatic MIR mode $\hat{a}_{\Omega,s}$ and a nonmonochromatic NIR mode $\sum_{i \in I} \tilde{\alpha}_i \hat{a}_{i,s}$ with the squeezing strength $\zeta$. In the following, we will discuss the conversion of Eq.~\eqref{eq:NL_action} into such an effective two-mode operator.

\subsection{The first order unitary}\label{sec:first_order}
In principle, it is possible to write the nonlinear unitary operator in Eq.~\eqref{eq:NL_action} in terms of discrete mode operators using a Schmidt decomposition. Thus, the time evolution in the nonlinear crystal can be understood as only acting on a countably infinite number of effective nonmonochromatic modes instead of a continuum of modes. However, a Schmidt decomposition is in general computationally exhausting and selecting a set of most significant modes participating in the interaction can be a challenging task. Instead, the first-order unitary $\hat{U}^{[1]} = \exp(\hat{S}^{[1]})$ with
\begin{equation}\label{eq:first_order_unitary}
	\hat{S}^{[1]} = \theta^{(1)}_{\tilde{\omega}}\left(\overline{\mathfrak{a}}_{\tilde{\omega}}\hat{\mathfrak{u}}_{\tilde{\omega}}^\dagger - \hc \right)
,\end{equation}
can be used in place of Eq.~\eqref{eq:NL_action}, as shown in Ref.~\cite{Onoe2022}. We will drop the index indicating the polarization from here on, since all operators in Eq.~\eqref{eq:first_order_unitary} act on the $s$-polarized modes. The broadband mode operator
\begin{equation}\label{eq:a_mode}
	\overline{\mathfrak{a}}_{\tilde{\omega}} = \frac{1}{\theta_{\tilde{\omega}}^{(1)}}[\hat{\mathfrak{u}}_{\tilde{\omega}},\hat{S}] = \int_{-\Lambda}^\Lambda f_{\tilde{\omega}}(\Omega) \e^{-\iu t_\text{p} \Omega} \hat{a}_\Omega \dd \Omega
,\end{equation}
with $f_{\tilde{\omega}}(\Omega) = (\theta_{\tilde{\omega}}^{(1)}\sqrt{\Delta \omega})^{-\frac{1}{2}} \int_{-\infty}^\infty \rect\left(\frac{\tilde{\omega}-\omega}{\Delta \omega}\right) S(\Omega,\omega) \dd \omega$ and $\theta_{\tilde{\omega}}^{(1)} = |[[\hat{S},\hat{\mathfrak{u}}_{\tilde{\omega}}^\dagger],[\hat{\mathfrak{u}}_{\tilde{\omega}},\hat{S}]]|^{\frac{1}{2}} \neq 0$, is not a pure annihilation or creation operator, since
\begin{equation}
	[\overline{\mathfrak{a}}_{\tilde{\omega}},\overline{\mathfrak{a}}_{\tilde{\omega}}^\dagger] = \int_{-\Lambda}^\Lambda \sign(\Omega) \abs{f_{\tilde{\omega}}(\Omega)}^2 \dd \Omega = \pm 1
.\end{equation}
For $\int_0^\Lambda \abs{f_{\tilde{\omega}}(\Omega)}^2 \dd \Omega > \int_{-\Lambda}^0 \abs{f_{\tilde{\omega}}(\Omega)}^2 \dd \Omega$ we will denote $\overline{\mathfrak{a}}_{\tilde{\omega}} = \hat{\mathfrak{a}}_{\tilde{\omega}}$ as an annihilation operator and for $\int_0^\Lambda \abs{f_{\tilde{\omega}}(\Omega)}^2 \dd \Omega < \int_{-\Lambda}^0 \abs{f_{\tilde{\omega}}(\Omega)}^2 \dd \Omega$ as a creation operator $\overline{\mathfrak{a}}_{\tilde{\omega}} = \hat{\mathfrak{a}}_{\tilde{\omega}}^\dagger$. In either case, this operator is in general different from the ordinary annihilation or creation operator, because $\overline{\mathfrak{a}}_{\tilde{\omega}}\ket{0} \neq 0$ and $\bra{0}\overline{\mathfrak{a}}_{\tilde{\omega}} \neq 0$. Only if $\int_{-\Lambda}^0 \abs{f_{\tilde{\omega}}(\Omega)}^2 \dd \Omega = 0$ or $\int_0^\Lambda \abs{f_{\tilde{\omega}}(\Omega)}^2 \dd \Omega = 0$, the operator $\overline{\mathfrak{a}}_{\tilde{\omega}}$ is actually an annihilation or creation operator of the (nonmonochromatic) Fock states.

\begin{figure}
    \includegraphics[width=0.485\textwidth]{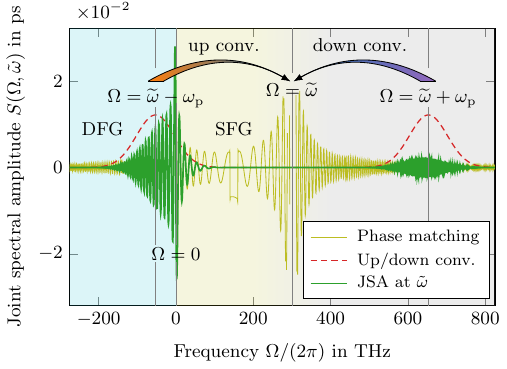}
	\caption{Joint spectral amplitude (JSA) $S(\Omega,\tilde{\omega})$ at the detected probe central frequency $\tilde{\omega}$, defining the mode function of $\overline{\mathfrak{a}}_{\tilde{\omega}}$ [cf. Eq.~\eqref{eq:out_mode_function}], as well as its two composing functions, the phase-matching function (solid light green line) $\zeta_{\Omega,\tilde{\omega}}$ and  $\alpha_\text{p}E_\text{p}(\tilde{\omega} - \Omega) + \conj{\alpha_\text{p}}\conj{E_\text{p}}(\Omega - \tilde{\omega})$ describing the up- and down-conversion of the $s$-polarized photons of frequency $\Omega$ (dashed red line). The negative frequencies $\Omega < 0$ of the up-conversion process correspond to the detected difference-frequency generation (DFG), $\tilde{\omega} = \omega_\text{p} - \abs{\Omega}$, while the positive frequencies $\Omega > 0$ correspond to sum-frequency generation (SFG), $\tilde{\omega} = \omega_\text{p} + \abs{\Omega}$. We assumed $\tilde{\omega} = \SI{300}{\tera\hertz}$, $\Delta \tilde{\omega} = \SI{1}{\tera\hertz}$, $\omega_\text{p} = \SI{350}{\tera\hertz}$, $\sigma_\text{p} = \SI{35}{\tera\hertz}$ and $L = \SI{100}{\micro\meter}$. The model for the refractive index is given in Eq.~\eqref{eq:refractive_index}. The phase-matching function has two maxima, one at $\Omega = 0$ and one at $\Omega = \tilde{\omega}$, while the pump contribution has  maxima at $\Omega = \tilde{\omega} - \omega_\text{p}$ and $\Omega = \tilde{\omega} + \omega_\text{p}$, indicated by the vertical lines.}
	\label{fig:MIR_spektrum}
\end{figure}

Because the filtered bandwidth $\Delta \omega$ is assumed to be narrow, the mode function can be approximated by 
\begin{align}
	&f_{\tilde{\omega}}(\Omega) \approx \frac{\sqrt{\Delta \omega}}{\theta_{\tilde{\omega}}^{(1)}} S(\Omega,\tilde{\omega})\e^{\iu t_\text{p} \Omega}, \label{eq:out_mode_function} \\
	&\theta_{\tilde{\omega}}^{(1)} \approx \sqrt{\Delta \tilde{\omega}}\abs{\int_{-\Lambda}^\Lambda \hspace{-2.5mm}\sign(\Omega)\abs{S(\tilde{\omega},\Omega)}^2\dd\Omega}^{1/2} \label{eq:theta_1}
.\end{align}
The spectrum of the MIR mode defined by Eq.~\eqref{eq:a_mode} is thus determined by the joint spectral amplitude $S(\Omega,\tilde{\omega})$ with one of the frequencies fixed at the central frequency $\tilde{\omega}$ of the probe, the detected frequency. An example of this spectrum can be seen in Fig.~\ref{fig:MIR_spektrum}. The detected up-conversion processes at $\Omega \sim \tilde{\omega} - \omega_\text{p}$ dominate the spectrum, since the down-conversion processes at $\Omega \sim \tilde{\omega} + \omega_\text{p}$ are suppressed by phase matching. The up-conversion can be separated into the detected DFG, $\tilde\omega = \omega_{\text{p}} - \abs{\Omega}$ for $\Omega < 0$ and SFG, $\tilde\omega = \omega_{\text{p}} + \abs{\Omega}$, for $\Omega > 0$. If the central frequency $\omega_{\text{p}}$ of the pump is higher than the central frequency of the probe, $\tilde{\omega} < \omega_{\text{p}}$, DFG corresponding to the squeezing contributions in Eq.~\eqref{eq:first_order_unitary} prevails and we can write $\overline{\mathfrak{a}}_{\tilde{\omega}} = \hat{\mathfrak{a}}_{\tilde{\omega}}^\dagger$.

\begin{figure}[ht]
    \includegraphics[width=0.486\textwidth]{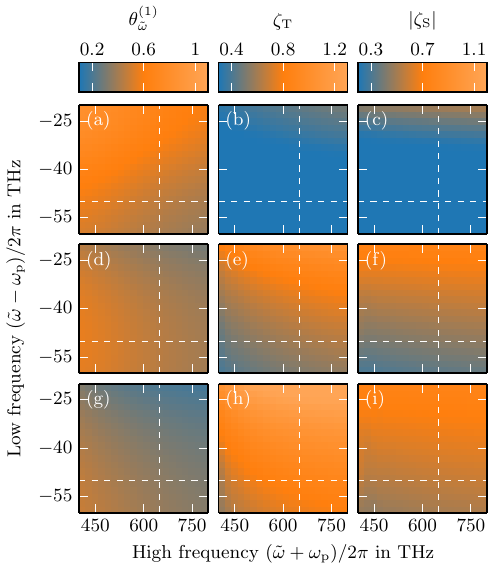}
	\caption{The three squeezing parameters determining the nonlinear unitary defined in Eq.~\eqref{eq:unitary_transform_nl} as a function of the up converted frequency $\tilde{\omega} - \omega_\text{p}$ and the down converted frequency $\tilde{\omega} + \omega_\text{p}$ [c.f. Fig.~\ref{fig:MIR_spektrum}]. Note that $\tilde\omega$ and $\omega_{\rm p}$ are both considered variables. The nonlinear crystal is of length $L = \SI{100}{\micro\meter}$ and the refractive index given in Eq.~\eqref{eq:refractive_index}. The filtered probe pulse bandwidth is chosen as $\Delta \tilde{\omega}/(2\pi) = \SI{1}{\tera\hertz}$ and the probe amplitude as $\alpha_\text{p} = 2\cdot 10^6$. The pump pulse bandwidth is $\sigma_\text{p}/(2\pi) = \SI{15}{\tera\hertz}$ for (a)-(c), $\sigma_\text{p}/(2\pi) = \SI{35}{\tera\hertz}$ for (d)-(f) and $\sigma_\text{p}/(2\pi) = \SI{50}{\tera\hertz}$ for (g)-(i). The white dashed lines indicate the point with $(\tilde{\omega} - \omega_\text{p})/(2\pi) = -\SI{50}{\tera\hertz}$ and $(\tilde{\omega} + \omega_\text{p})/(2\pi) = \SI{650}{\tera\hertz}$, which are the values used in our example for time-domain quantum state tomography in Sec.~\ref{sec:tomography}.}
	\label{fig:squeezing_params}
\end{figure}

From now on, we focus on the case $\overline{\mathfrak{a}}_{\tilde{\omega}} = \hat{\mathfrak{a}}_{\tilde{\omega}}^\dagger$, which we will refer to as the squeezing regime. In this case, it is possible to decompose the operator as
\begin{align*}
	\overline{\mathfrak{a}}_{\tilde{\omega}}^\dagger = \hat{\mathfrak{a}}_{\tilde{\omega}} =& \cosh(\theta)\hat{\mathfrak{a}}_{\text{SA}} + \sinh(\theta)\sin(\theta_\perp)\e^{-\iu\Phi_\perp}\hat{\mathfrak{a}}_{\text{SA}}^\dagger \\
	&+ \sinh(\theta)\cos(\theta_\perp)\hat{\mathfrak{a}}_{\text{TH}}^\dagger \numberthis \label{eq:mode_op_decomposition}
,\end{align*}
with $\theta = \arccosh(\sqrt{\int_{-\Lambda}^0 \abs{f_{\tilde{\omega}}(\Omega)}^2 \dd \Omega})$ and $\sin(\theta_\perp)\e^{\iu\Phi_\perp} \allowbreak= \csch(\theta)\sech(\theta)\int_0^\Lambda f_{\tilde{\omega}}(\Omega)f_{\tilde{\omega}}(-\Omega)\dd \Omega$, as well as the two discretized (pure) annihilation operators $\hat{\mathfrak{a}}_{\text{SA/TH}} = \int_0^\Lambda f_{\text{SA/TH}}(\Omega)\e^{-\iu t_\text{p}\Omega}\hat{a}_\Omega \dd \Omega$ with
\begin{align*}
	f_{\text{SA}}(\Omega) =& \sech(\theta) \conj{f}_{\tilde{\omega}}(-\Omega), \numberthis \\
	f_{\text{TH}}(\Omega) =& \sec(\theta_\perp) \csch(\theta) f_{\tilde{\omega}}(\Omega) \\
	&- \tan(\theta_\perp)\e^{\iu\Phi_\perp}f_{\text{SA}}(\Omega) \numberthis
.\end{align*}
As will become clear later, the DFG contribution and thus the operator $\hat{\mathfrak{a}}_{\text{SA}}$ corresponds to the part of the MIR spectrum sampled by the electro-optic sampling setup in Fig.~\ref{fig:setup}, while $\hat{\mathfrak{a}}_{\text{TH}}$ represents the thermalized modes, which are unsampled but entangled through the nonlinear interaction to the sampled modes. By tracing over the entangled modes, the state of the sampled modes becomes partially mixed, resulting in an increased von Neumann entropy. Thus, the mode annihilated by $\hat{\mathfrak{a}}_{\text{TH}}$ thermalizes the state of the sampled modes \cite{Onoe2022}. The two operators commute by construction $\comm{\hat{\mathfrak{a}}_{\text{SA}}}{\hat{\mathfrak{a}}_{\text{TH}}} = 0$ and because of the way the operators are constructed, the thermalized mode depends on the sampled mode. If the phase difference between the sampled mode and the modes contributing to the SFG is a multiple of $\pi$, $\Phi_\perp = n\pi$, the sampled frequencies attenuate the thermalized mode, as will become relevant in Sec.~\ref{sec:tomography}. The decomposition in Eq.~\eqref{eq:mode_op_decomposition} can be expressed using a unitary single-mode squeezing operator $\hat{S}\hat{\mathfrak{a}}_{\text{SA}}\hat{S}^\dagger = \mu_{\text{S}}\hat{\mathfrak{a}}_{\text{SA}} + \nu_{\text{S}} \hat{\mathfrak{a}}_{\text{SA}}^\dagger$ and two-mode squeezing operator $\hat{T}\hat{\mathfrak{a}}_{\text{SA}}\hat{T}^\dagger = \mu_{\text{T}}\hat{\mathfrak{a}}_{\text{SA}} + \nu_{\text{T}} \hat{\mathfrak{a}}_{\text{TH}}^\dagger$
with $\mu_{\text{S/T}} = \cosh(\abs{\zeta_{\text{S/T}}})$, $\nu_{\text{S/T}} = \exp[\iu\arg(\zeta_{\text{S/T}})]\sinh(\abs{\zeta_{\text{S/T}}})$. In order to write
\begin{equation}\label{eq:unitary_decomp}
	\hat{\mathfrak{a}}_{\tilde{\omega}} = \hat{S}\hat{T}\hat{\mathfrak{a}}_{\text{SA}}\hat{T}^\dagger\hat{S}^\dagger
,\end{equation}
we have to choose 
\begin{align}
    \zeta_{\text{S}} &= \e^{\iu\Phi_\perp}\arccosh\left\{[1-\tanh^2(\theta)\sin^2(\theta_\perp)]^{-\frac{1}{2}}\right\} \label{eq:single_mode_squeezing_param}, \\
    \zeta_{\text{T}} &= \arccosh\left\{[\cosh^2(\theta)-\sinh^2(\theta)\sin^2(\theta_\perp)]^{\frac{1}{2}}\right\} \label{eq:two_mode_squeezing_param}
.\end{align}
The squeezing parameters $\theta_{\tilde{\omega}}^{(1)}$, $\abs{\zeta_{\text{S}}}$ and $\zeta_{\text{T}}$ are numerically evaluated as a function of the up-converted frequency $\tilde{\omega} - \omega_\text{p}$ and the down-converted frequency $\tilde{\omega} + \omega_\text{p}$, in Fig.~\ref{fig:squeezing_params}. If the maximum of the pump shape corresponding to the up-conversion process $\tilde{\omega} - \omega_\text{p}$ is shifted towards the origin or if the bandwidth of the pump $\sigma_\text{p}$ is increased, the single- and two-mode squeezing parameters increase, since more SFG is involved in the nonlinear interaction, agreeing with the intuitive picture developed in Fig.~\ref{fig:MIR_spektrum}.
Inserting Eq.~\eqref{eq:unitary_decomp} in the first-order unitary defined by Eq.~\eqref{eq:first_order_unitary} results in 
\begin{equation}\label{eq:unitary_transform_nl}
	\hat{U}^{[1]} = \hat{S}\hat{T}\hat{U}_\text{NL}^\prime\hat{T}^\dagger\hat{S}^\dagger
,\end{equation}
with the effective two-mode squeezing operator
\begin{equation}\label{eq:NL_effective}
	\hat{U}_\text{NL}^\prime = \exp[-\theta_{\tilde{\omega}}^{(1)}\left(\hat{\mathfrak{a}}_{\text{SA}}\sum_{i \in I}\tilde{\alpha}_i\hat{\mathfrak{u}}_{\tilde{\omega}_i} - \hc\right)]
,\end{equation}
similar to the operator in Eq.~\eqref{eq:U_effective}, used in Ref.~\cite{Hubenschmid2022}. Utilizing the unitary transformation of the quadratic unitary operator in Eq.~\eqref{eq:unitary_transform_nl}, the total time evolution can be expressed as
\begin{equation}\label{eq:U_decomp}
	\hat{U} \approx \hat{S}\hat{T}\hat{U}_{\text{eff}}\hat{T}^\dagger\hat{S}^\dagger
,\end{equation}
where
\begin{equation}\label{eq:U_effective}
	\hat{U}_{\text{eff}} = \hat{U}_{\text{WP}} \hat{D}_{\tilde{\omega}}(\vec{\beta})\hat{U}_\text{NL}^\prime
\end{equation}
is the total, transformed time-evolution operator. The unitary operator $\hat{U}_{\text{eff}}$ is mathematically equivalent to the time evolution used in Ref.~\cite{Hubenschmid2022} and thus, by transforming the MIR state using the squeezing operators $\hat{S}$ and $\hat{T}$, one can apply the results from Ref.~\cite{Hubenschmid2022}, as will be done in the following section.

\section{The difference-count probability distribution}\label{sec:result}
The $s$-quasiprobability distributions $\rho(z;s)$ parameterized by a single (real) number $s$ are a representation of a (possibly mixed) quantum state equivalent to the density operator $\hat{\rho}$, in the sense that all observable quantities calculated from either agree. The two representations can be related to one another via the $s$-characteristic function $\chi(\beta;s) = \exp(s\abs{\beta}^2/2)\tr[\hat{D}(\beta)\hat{\rho}]$, which is the Fourier transform of the $s$-quasiprobability distribution $\rho(z;s) = \pi^{-1} \int \exp(-2\iu\Im[\beta\conj{z}])\chi(\beta;s)\dd^2 \beta$. For $s=-1,0,1$, the characteristic function corresponds to the expectation value of the antinormally, symmetrically and normally ordered displacement operator and the respective $s$-quasiprobability distributions are the Husimi, Wigner and Glauber-Sudarshan distributions \cite{Cahill1969,Carmichael1999,Vogel2006}. For the electro-optic sampling of a single monochromatic mode, we have recently shown that the probability distribution of the measured photon-number difference is related to a $s$-quasiprobability distribution of the sampled quantum state \cite{Hubenschmid2022}.
In the case of electro-optic samping with an ultrabroadband pump using the setup described in the previous section, the probability to measure the photon-number differences $\{\Delta n_i\}$ given the MIR modes, centered at $\tilde{\Omega}$, are in the initial broadband state $\hat{\rho}_{\tilde{\Omega}}$ reads
\begin{equation}\label{eq:prob_dist_def}
    p(\{\Delta n_i\}) = \tr(\hat{P}_{\{\Delta n_i\}} \hat{U}\hat{\rho}_{\tilde{\Omega}} \otimes \ket{0}_\text{NIR}\prescript{}{\text{NIR}}{\bra{0}} \hat{U}^\dagger)
.\end{equation}
We assume that the (ultrabroadband) NIR modes (after the removal of the $z$-polarized pump) are in the ground state $\ket{0}_{\text{NIR}}$ and the combined state of MIR and NIR modes is evolved using the unitary time-evolution operator $\hat{U}$ from Eq.~\eqref{eq:U_decomp} until a projective measurement, described by the projectors $\hat{P}_{\{\Delta n_i\}}$ defined in Eq.~\eqref{eq:projectors}, is performed. Inserting Eq.~\eqref{eq:U_decomp} into Eq.~\eqref{eq:prob_dist_def} yields
\begin{align*}
	p(\{\Delta n_i\}) \approx& \tr\left[\hat{P}_{\{\Delta n_i\}}\hat{S}\hat{T}\hat{U}_{\text{eff}}\hat{T}^\dagger\hat{S}^\dagger \hat{\rho}_{\tilde{\Omega}}\right. \\ 
	&\left. \otimes \ket{0}_\text{NIR}\prescript{}{\text{NIR}}{\bra{0}} \hat{S}\hat{T}\hat{U}_{\text{eff}}^\dagger\hat{T}^\dagger\hat{S}^\dagger\right] \numberthis
.\end{align*}
The projectors $\hat{P}_{\{\Delta n_i\}}$ defined in Eq.~\eqref{eq:projectors} and the squeezing operators $\hat{S}$, $\hat{T}$ act on different frequency bands, and therefore, they commute. Thus, by defining a transformed and reduced density operator (in which $\tr_{\text{r}}$ is the trace over the modes not involved in the nonlinear interaction)
\begin{equation}\label{eq:transformed_state}
    \hat{\tilde{\rho}}_{\text{SA}} = \tr_{\text{r}}\left[\hat{S}^\dagger\hat{T}^\dagger \hat{\rho}_{\tilde{\Omega}}\hat{T}\hat{S}\right]
\end{equation}
solely of the sampled modes we can recast the count-probability distribution into
\begin{align*}\label{eq:prob_dist_cont}
	p(\{\Delta n_i\}) \approx& \tr_\text{NIR}\Bigg\{\tr_{\text{SA}}\Bigg[\hat{P}_{\{\Delta n_i\}} \\
	&\times \hat{U}_{\text{eff}} \hat{\tilde{\rho}}_{\text{SA}} \otimes \ket{0}_\text{NIR}\prescript{}{\text{NIR}}{\bra{0}} \hat{U}_{\text{eff}}^\dagger\Bigg]\Bigg\} \numberthis
.\end{align*}
Eq.~\eqref{eq:prob_dist_cont} only involves the (nonmonochromatic) sampled MIR mode and the NIR modes and allows one to apply the formalism developed in Ref.~\cite{Hubenschmid2022}. Thus, we can relate the count-probability distribution to the $\tilde{s}$-quasiprobability distribution $\tilde{\rho}_\text{SA}(z;s)$ of the transformed state in Eq.~\eqref{eq:transformed_state} through
\begin{equation}\label{eq:prob_dist}
	p(\{\Delta n_i\}) \approx N\tilde{\rho}_\text{SA}(z(\{\Delta n_i\});\tilde{s})
.\end{equation}
The quasiprobability distribution with parameter $\tilde{s} = 1 - 2\coth^2(|\theta_{\tilde{\omega}}^{(1)}|)$ depending on the squeezing parameter $\theta_{\tilde{\omega}}^{(1)}$ of the nonlinear interaction [cf. Eq.~\eqref{eq:first_order_unitary}] is renormalized by $N = \csch^2(|\theta_{\tilde{\omega}}^{(1)}|)/2$ and related to the discrete photon-count differences via its argument
\begin{equation}
    z(\{\Delta n_i\}) = \frac{1}{\sqrt{2}}\csch(|\zeta|)\left(\frac{\Delta n_{\text{X}}}{\beta_{\text{X}}} + \iu \frac{\Delta n_{\text{Y}}}{\beta_{\text{Y}}}\right)
,\end{equation}
with amplitudes $\beta_i$ of the post-filtered probes. See Sec.~\ref{sec:heuristics} for an intuitive explanation of this result. Eq.~\eqref{eq:prob_dist} holds only for $\abs{\tilde{\alpha}_{\text{X}}}^2 = \abs{\tilde{\alpha}_{\text{Y}}}^2 = 1/2$. However, the result presented here can easily be generalized to measurements of multiple $\hat{X}$ and $\hat{Y}$ quadratures with different $\alpha_i$ by applying the full result of Ref.~\cite{Hubenschmid2022}. The simplified count-probabilty distribution suffices for our purpose. The nonmonochromatic $s$-quasiprobability distribution $\hat{\tilde{\rho}}_{\text{SA}}$ of the transformed state can be related to the two-mode characteristic function $\chi_{\text{TM}}(\beta_{\text{SA}},\beta_{\text{TH}}) = \tr[\hat{D}_{\text{SA}}(\beta_{\text{SA}})\hat{D}_{\text{TH}}(\beta_{\text{TH}})\hat{\rho}_{\tilde{\Omega}}]$, defined as the expectation value of the displacement operators acting on the modes SA and TH, by
\begin{align*}\label{eq:cont_qpd}
	\tilde{\rho}_{\text{SA}}(z;s) =& \frac{1}{\pi} \int \chi_{\text{SA}}(\beta;s) \e^{-2\iu\Im(\beta\conj{z})} \dd^2 \beta \\
	=&\frac{1}{\pi} \int \chi_{\text{TM}}\left[\mu_{\text{T}}(\mu_{\text{S}}\beta - \nu_{\text{S}}\conj{\beta}),-\nu_{\text{T}}\conj{\beta}\right] \\
	&\times \e^{-2\iu\Im(\beta\conj{z}) + s\abs{\beta}^2/2} \dd^2 \beta \numberthis
,\end{align*}
as shown in Appendix~\ref{a:two_mode_char}. The transformed $s$-quasiprobability distributions for broadband coherent states, cat states and the squeezed vacuum can be taken from Appendix~\ref{a:quasi_prob}.

\begin{figure}
    \includegraphics[width=0.48\textwidth]{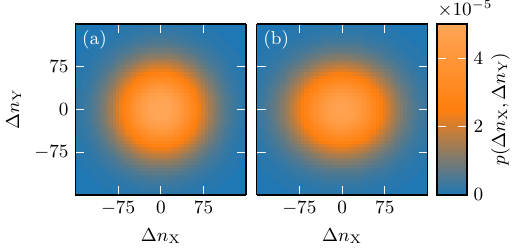}
	\caption{Count-probability distribution  [Eq.~\eqref{eq:prob_dist}] for a broadband MIR vacuum state with $\tilde{\Omega}/(2\pi) = \SI{25}{\tera\hertz}$, $\sigma_{\Omega}/(2\pi) = \SI{5}{\tera\hertz}$. For the filtered probe, we assume $\tilde{\omega}/(2\pi) = \SI{300}{\tera\hertz}$, $\Delta \omega/(2\pi) = \SI{1}{\tera\hertz}$ and $\beta = 50$. We consider a crystal of length $L = \SI{100}{\micro\meter}$. The parameters describing the pump are $\alpha_\text{p} = 2 \cdot 10^6$, $t_\text{p} = \SI{0}{\pico\second}$, $\omega_\text{p}/(2\pi) = \SI{350}{\tera\hertz}$. (a) The bandwidth of the pump is $\sigma_\text{p}/(2\pi) = \SI{15}{\tera\hertz}$ and difference-frequency generation dominates the nonlinear interaction. (b) The pump bandwidth is $\sigma_\text{p}/(2\pi) = \SI{50}{\tera\hertz}$ and the amount of sum-frequency generation to the nonlinear interaction is enhanced compared to the case in (a), resulting in some thermalization of the sampled quasiprobabilty distribution.}
	\label{fig:sfg_effect}
\end{figure}

The transformed quasiprobability distribution in Eq.~\eqref{eq:cont_qpd} turns into the quasiprobability distribution of the sampled state $\hat{\rho}_{\text{SA}} = \tr_{\text{r}}(\hat{\rho}_{\tilde{\Omega}})$, if only DFG contributes to the nonlinear interaction, i.e., $\zeta_{\text{S}}, \zeta_{\text{T}} = 0$. Increasing the SFG will add some contribution from the thermalized mode to the transformed state $\tilde{\rho}_{\text{SA}}$. If the MIR mode is in the vacuum state and if the  bandwidth of the pump is assumed to be $\sigma_{\text{p}}/(2\pi) = \SI{15}{\tera\hertz}$, DFG predominates and the difference-count probability distribution follows the quasiprobability distribution of the vacuum, as can be seen in Fig.~\ref{fig:sfg_effect} (a). However, if the  bandwidth of the pump is assumed to be $\sigma_{\text{p}}/(2\pi) = \SI{50}{\tera\hertz}$, SFG contributes stronger to the nonlinear interaction and the difference-count probability distribution is squeezed due to the effect of the squeezing operators in Eq.~\eqref{eq:transformed_state} on the sampled state, visible in Fig.~\ref{fig:sfg_effect} (b). This observation is also supported by the ensemble calculation in Appendix \ref{a:therm}. The variance of the difference counts $\Delta n_i$ has two contributions, one dependent on the MIR state and one only dependent on the parameter of the nonlinear interaction $\tilde{s} = 1 - 2\coth^2(|\theta_{\tilde{\omega}}^{(1)}|)$. The contribution from the latter can be reduced by increasing the squeezing parameter $\theta_{\tilde{\omega}}^{(1)}$, which can be achieved by increasing the amplitude $\alpha_\text{p}$ of the pump or by shifting the central frequencies $\tilde{\omega}$, $\omega_\text{p}$ towards increasing $\theta_{\tilde{\omega}}^{(1)}$ according to Fig.~\ref{fig:squeezing_params}. The state-dependent contribution can be expressed in terms of the variance of the transformed sampled state in the MIR and hence includes contributions from the thermalized mode. For a coherent MIR input state $\hat{\rho}_{\tilde{\Omega}}$ (including the vacuum as a limit case), the main result from Ref.~\cite{Onoe2022} in the squeezing regime can be reproduced and generalized. Eq.~\eqref{eq:var_coh} shows that in the squeezing regime, i.e., $\overline{\mathfrak{a}}_{\tilde{\omega}} = \hat{\mathfrak{a}}_{\tilde{\omega}}^\dagger$, thermalization leads to an increase of the variance, which we will call thermalization noise from here on. The thermalization noise can be mitigated by filtering below the pump central frequency, $\tilde{\omega} \leq \omega_\text{p}$, as we discuss in the next section.

\begin{figure*}
    \includegraphics[width=1\textwidth]{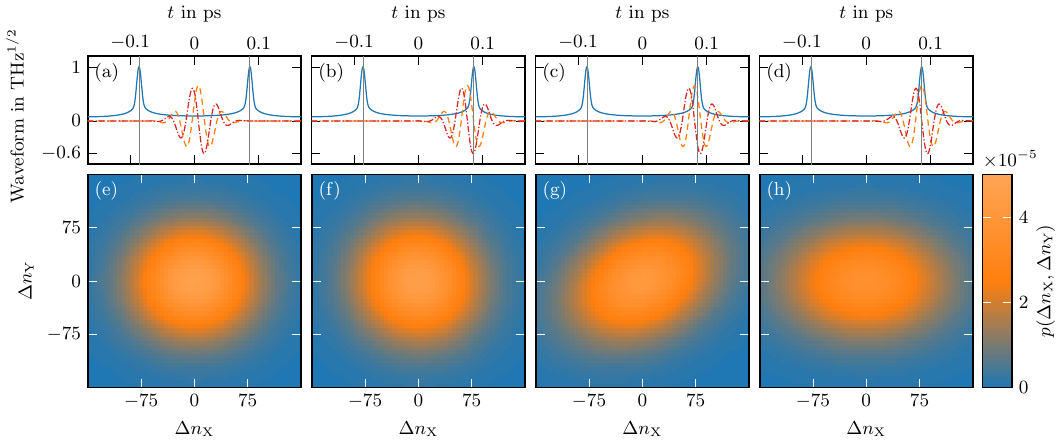}
	\caption{(a)-(d) Envelope of the gating function $\abs{\mathcal{F}[f_{\text{SA}}](t)}$ (solid blue line) and the waveform of the MIR $\mathcal{F}[f_{\tilde{\Omega}}](t)$ (dashed orange line for the real part and dash-dotted red line for the imaginary part) in the time domain. The sampled mode has two maxima indicated by the vertical lines located one period $\eta_\text{c} = \frac{L}{2c}[n_\text{g}(\tilde{\omega}) - n(0)]$ of the (approximated) phase-matching function  away from the origin. The carrier envelope phase of the MIR $t_{\tilde{\Omega}} = \eta_\text{c} - 2\pi t_\text{step} / \tilde{\Omega}$ is varied in (a)-(d) by $t_\text{step} = 2,\frac{3}{4},\frac{1}{8},0$. The pump pulse is a coherent state with amplitude $\alpha_\text{p} = 2\cdot 10^6$ and of a Gaussian shape defined by Eq.~\eqref{eq:probe_spec} centered at $\omega_\text{p}/(2\pi) = \SI{350}{\tera\hertz}$ and of bandwidth $\sigma_\text{p}/(2\pi) = \SI{35}{\tera\hertz}$ with a fixed carrier envelope phase $t_\text{p} = \SI{0}{\second}$, while the MIR is a Gaussian centered at $\tilde{\Omega}/(2\pi) = \SI{25}{\tera\hertz}$ and of bandwidth $\sigma_{\tilde{\Omega}}/(2\pi) = \SI{5}{\tera\hertz}$. (e)-(h) Example of the count-probability distribution in Eq.~\eqref{eq:prob_dist} for a squeezed vacuum generated by the operator in Eq.~\eqref{eq:mir_squeezing_op} with $\zeta_{\tilde{\Omega}} = 1.5$. Eq.~\eqref{eq:cont_qpd} shows the relation between the count-probability distributions over $\{\Delta n_{\text{X}}, \Delta n_{\text{Y}}\}$ and the quasiprobability distribution of the sampled quantum state, associating $\Delta n_{\text{X}}$ with $\hat{X}$-quadrature and $\Delta n_{\text{Y}}$ with $\hat{Y}$-quadrature measurements. The filtered probe pulse is of amplitude $\beta = \sqrt{2} \times 10$ and of bandwidth $\Delta \omega/(2\pi) = \SI{1}{\tera\hertz}$ at $\tilde{\omega}/(2\pi) = \SI{300}{\tera\hertz}$. The crystal is assumed to be of length $L = \SI{100}{\micro\meter}$ and made of zinc telluride with the refractive index of Fig.~\ref{fig:refrative_index}.} 
	\label{fig:example}
\end{figure*}

If the broadband MIR mode is in some arbitrary state $\hat{\rho}_{\tilde{\Omega}}(\hat{\mathfrak{a}}_{\tilde{\Omega}})$ defined in terms of a single nonmonochromatic mode operator $\hat{\mathfrak{a}}_{\tilde{\Omega}} = \int_0^\infty f_{\tilde{\Omega}}(\Omega) \e^{-\iu t_{\tilde{\Omega}}\Omega} \hat{a}_\Omega \dd\Omega$, the two-mode characteristic function can be calculated using the decomposition
\begin{equation}\label{eq:mir_mode_op_decomposition}
	\hat{\mathfrak{a}}_{\tilde{\Omega}} = A_{\text{SA}}(\Delta t)\hat{\mathfrak{a}}_{\text{SA}} + A_{\text{TH}}(\Delta t)\hat{\mathfrak{a}}_{\text{TH}} + A_\text{UN}(\Delta t)\hat{\mathfrak{a}}_\text{UN}
,\end{equation}
where the coefficients $A_{\text{x}}(\Delta t) = [\hat{\mathfrak{a}}_{\tilde{\Omega}}, \hat{\mathfrak{a}}_\text{x}^\dagger]$ ($\text{x} = \text{SA,TH,UN}$) 
quantify the amount the sampled, thermalized, and unsampled mode contribute to the MIR mode. The newly introduced operator $\hat{\mathfrak{a}}_\text{UN}$ of the unsampled and uncorrelated modes, leaving the electro-optic signal unaffected, has to be differentiated from the thermalized mode, which is also unsampled, but correlated to the sampled mode, therefore altering the measurement. All operators on the right-hand side of Eq.~\eqref{eq:mir_mode_op_decomposition} commute. The coefficients $A_{\text{x}}(\Delta t) $ can be expressed as the convolution of the Fourier-transformed mode functions $\mathcal{F}[f_{\tilde{\Omega}}](t)$ and the gating function $\mathcal{F}[\conj{f}_{\text{x}}](t)$, dependent on the time delay $\Delta t = t_{\tilde{\Omega}} - t_{\text{p}}$ between the sampled MIR pulse and the NIR pump pulse according to
\begin{equation}\label{eq:Convolution}
	A_{\text{x}}(\Delta t) \approx  (\mathcal{F}[f_{\tilde{\Omega}}] \ast \mathcal{F}[\conj{f}_{\text{x}}])(\Delta t).
\end{equation}
Therefore, the coefficients can be understood as the waveform of the MIR pulse $\mathcal{F}(f_{\tilde{\Omega}})$ averaged over a time window defined by $\mathcal{F}(f_{\text{x}})$. If the waveform is zero on average and thus $|A_{\text{UN}}(\Delta t)|=1$, the transformed $s$-quasiprobability distribution corresponds to the (possibly squeezed) vacuum, shown in Fig.~\ref{fig:example} (a). The key idea to understand our approach to time-domain quantum state tomography is the following: As the time delay $\Delta t$ is varied, the time window is shifted and different parts of the MIR waveform are averaged, which can increase the contributions from the MIR state to the transformed quasiprobability distribution $\tilde{\rho}_{\text{SA}}(z;s)$ depending on the average waveform, thus dynamically sampling the MIR state. This can be seen in Fig.~\ref{fig:example} (a)-(d) for the squeezed vacuum in the MIR. The shorter the NIR pump, the shorter is the time window $\mathcal{F}(f_{\text{SA}})$ the MIR pulse is averaged over. However, a shorter pump pulse will lead to an increased bandwidth $\sigma_{\text{p}}$ of the pump, because of the Gabor limit \cite{Gabor1946}. The broad bandwidth leads to an increase of the thermalization noise, which can be mitigated by filtering, as will be discussed in the next section. This is the principle idea behind time-domain optical tomography, as proposed in this work.

\section{Optical time-domain quantum tomography}\label{sec:tomography}
A phase-sensitive measurement typically requires the comparison of the signal to a local oscillator. For pulse-based homodyne detection, the sampled pulsed modes and the local oscillator pulse are mixed in a beam splitter, while for electro-optic sampling they interact in a nonlinear crystal. The mixing unavoidably leads to an averaging of the detected modes over the duration (and oscillations) of the local oscillator pulse. The goal of optical time-domain quantum tomography is to reconstruct the quantum state of a system, namely the MIR state, such that the averaging is taken over a time interval below the duration of a single cycle of the MIR-pulse central frequency (i.e., subcycle), granting access to the dynamics of quantum systems. The reconstruction of the sampled MIR state and its dynamics using the quantum-electro-optic-sampling setup proposed in Sec. \ref{sec:theoretical_description} is based on the relation between the photon-count probability distribution of the setup and the transformed $\tilde{s}$-quasiprobability distribution in Eq.~\eqref{eq:prob_dist}. For each fixed time delay $\Delta t$ between the NIR pump pulse and the sampled MIR pulsed mode, the transformed MIR $\tilde{s}$-quasiprobability distribution can be reconstructed from sampled data using the relation in Eq.~\eqref{eq:prob_dist}. Performing the reconstruction for various time delays $\Delta t$ gives access to the time dependence of the transformed MIR state and thus to the coefficients $A_\text{SA/TH}(\Delta t)$ defined through Eq.~\eqref{eq:mir_mode_op_decomposition}.

The remaining task is the reconstruction of the waveform of the MIR mode from the transformed quasiprobability distribution. As discussed in the preceding section, the sampled quadratures follow the MIR waveform if an ultrabroadband pump pulse is used. To make this argument more quantitative we consider the example of a MIR pulse with a Gaussian-mode function $f_{\tilde{\Omega}}(\Omega) = N_{\tilde{\Omega}}\sqrt{\Omega} \exp[-(\Omega - \tilde{\Omega})^2/(2\sigma_{\tilde{\Omega}})^2]$ normalized by $N_{\tilde{\Omega}}$. In this case, the coefficients 
\begin{equation}\label{eq:A_D_approx}
	A_{\text{SA}}(\Delta t) \approx\hspace{-0.6mm} -\sech(\theta) \frac{\sqrt{\Delta \tilde{\omega}}}{\theta_{\tilde{\omega}}^{(1)}} \abs{S_\text{c}} \e^{\iu\Phi_\perp/2}\hspace{-1.1mm}\sqrt{\frac{\pi}{n(0)}} N_{\tilde{\Omega}} \bar{\sigma} a_{(-)}(\Delta t) 
\end{equation}
and
\begin{align*}\label{eq:A_Th_approx}
	A_{\text{TH}}(\Delta t) \approx& -\frac{\sqrt{\Delta \tilde{\omega}}}{\theta_{\tilde{\omega}}^{(1)}} \abs{S_\text{c}} \e^{\iu\Phi_\perp/2}\sqrt{\frac{\pi}{n(0)}} N_{\tilde{\Omega}} \bar{\sigma} \\
	&\times \Big[\csch(\theta)\sec(\theta_{\perp})\e^{-\iu \Phi_\perp} a_{(+)}(\Delta t) \\
    &- \sech(\theta)\tan(\theta_\perp)\e^{\iu\Phi_\perp} a_{(-)}(\Delta t)\Big] \numberthis
\end{align*}
are approximately given in terms of Gaussian profiles 
\begin{align*}\label{eq:measured_pulse}
	a_{(\mp)}(\Delta t) &= \exp[-\frac{(\omega_\text{p} - \tilde{\omega})^2}{4\sigma_\text{p}^2} - \frac{\tilde{\Omega}^2}{4\sigma_{\tilde{\Omega}}^2} + \frac{\bar{\Omega}_{(\mp)}^2}{4\bar{\sigma}^2}] \\
	&\times \Big\{\exp[-(\Delta t - \eta_\text{c})^2\bar{\sigma}^2 - \iu(\Delta t - \eta_\text{c})\bar{\Omega}_{(\mp)}] \\
	&- \exp[-(\Delta t + \eta_\text{c})^2\bar{\sigma}^2 - \iu(\Delta t + \eta_\text{c})\bar{\Omega}_{(\mp)}]\Big\} \numberthis
,\end{align*}
with the inverse-variance average $\bar{\sigma} = \sigma_\text{p}\sigma_{\tilde{\Omega}}/\sqrt{\sigma_\text{p}^2 + \sigma_{\tilde{\Omega}}^2}$ and the weighted average central frequency $\bar{\Omega}_{(\mp)} = \mp (\tilde{\omega} - \omega_\text{p})\frac{\bar{\sigma}^2}{\sigma_\text{p}^2} + \tilde{\Omega}\frac{\bar{\sigma}^2}{\sigma_{\tilde{\Omega}}^2}$. The coefficients in Eq.~\eqref{eq:A_D_approx} and \eqref{eq:A_Th_approx} are obtained after approximating the phase-matching function [c.f. Eq.~\eqref{eq:phase_matching}] by $S_{\text{c}}\sinc(\eta_{\text{c}}\Omega)$, with constant $S_\text{c}$ given by Eq.~\eqref{eq:S_c} and $\eta_\text{c} = \frac{L}{2c}[n_\text{g}(\tilde{\omega}) - n(0)]$ dependent on the group refractive index $n_{\text{g}}(\tilde{\omega})$. The approximation is valid as long as the refractive index of the nonlinear crystal is flat in the MIR frequency band or in other words dispersion is low (see Appendix~\ref{a:coefficients} for details). The coefficients from Eq.~\eqref{eq:A_D_approx} and \eqref{eq:A_Th_approx} are compared to the MIR waveform in Fig.~\ref{fig:coefficients} as functions of the time delay $\Delta t$ between the NIR pump pulse and the sampled MIR mode for various pump bandwidths $\sigma_\text{p}$. Increasing the pump bandwidth improves the matching between the coefficient of the sampled mode $A_{\text{SA}}(\Delta t)$ and MIR mode profile. However, as further explained below, there are six effects due to the convolution of the MIR waveform with the gating function in Eq.~\eqref{eq:Convolution} that detrimentally affect this matching.

\begin{enumerate}
\item \emph{Desynchronization.}---The most prominent effect is desynchronization between $A_{\text{SA}}(\Delta t)$ and the MIR waveform by time shifts $\pm\eta_\text{c}$ due to the oscillation of the phase-matching in the frequency domain. If the shift is large enough (e.g., $\eta_{\text{c}} \gg \sigma_{\tilde{\Omega}}$), the two contributions at $-\eta_\text{c}$ and $\eta_\text{c}$ do not interfere and the shift can be compensated by adjusting $\Delta t$. The shift increases with the length of the nonlinear crystal and with $\tilde{\omega}$. A proper selection of these parameters therefore allows for some degree of control over the shift. Nonetheless, for a fixed pump central frequency $\omega_{\text{p}}$, the increase in $\tilde{\omega}$ makes the measurement weaker [i.e., decreases $\theta_{\tilde{\omega}}^{(1)}$] and enhances thermalization by increasing $\zeta_{\text{T}}$, what can be counteracted by increasing $\omega_{\text{p}}$ alongside $\tilde{\omega}$. We highlight suitable values by the white dashed lines in Fig.~\ref{fig:squeezing_params}.

\item \emph{Phase shifting.}---The real part of $A_{\text{SA}}(\Delta t)$ corresponds to the imaginary part of the MIR waveform and vice versa, as can be seen in Fig.~\ref{fig:coefficients}. This originates from the imaginary prefactor in the electric field and carries over to the gating function $\mathcal{F}[\conj{f}_{\text{x}}](t)$. This effect is absent in Fig.~\ref{fig:coeff_max}, since the interference of the two Gaussians at $\pm\eta_{\text{c}}$, composing the gating function $\mathcal{F}[\conj{f}_{\text{x}}](t)$, leads to an additional imaginary prefactor.

\item \emph{Spectral weighting.}---The low-frequency end of the MIR spectrum is overestimated. This skewed spectral weighting results from the $1/\sqrt{\Omega}$-dependence of the function $f_{\text{SA}}(\Omega)$ defining the sampled mode. This effect is, however, negligible as long as the central frequency of the sampled MIR mode is large compared to its own bandwidth $\tilde{\Omega} \gg \sigma_{\tilde{\Omega}}$. This effect can be seen by comparing Fig.~\ref{fig:coefficients} (b) and (c): in (b), $A_{\text{SA}}(\Delta t)$ seems to follow the MIR waveform better than in (c), even though in the latter the pump is shorter. The underestimation can be accounted for by reintroducing the factor $\sqrt{\Omega}$ in post-processing. 

\item \emph{Temporal averaging.}---The MIR waveform is averaged over the time window defined by the gating function. For a Gaussian MIR pulsed mode with central frequency $\tilde{\Omega}$ and variance $\sigma_{\tilde{\Omega}}$, the averaging will lead to the reconstruction of a Gaussian with an effective central frequency $\bar{\Omega}_{(-)}$ and variance $\bar{\sigma}$. When considering the limiting case of an infinitely broadband pump pulse, $\lim_{\sigma_\text{p} \to \infty} \bar{\sigma} = \sigma_{\tilde{\Omega}}$ and $\lim_{\sigma_\text{p} \to \infty} \bar{\Omega}_{(\mp)} = \tilde{\Omega}$, the reconstructed parameters tend to the actual MIR parameters, as shown in Fig.~\ref{fig:coefficients}. This (unphysical) limit case would correspond to a probe with amplitude described by a delta distribution over time, which "averages" over a single point of the sampled mode (the instant at which they intersect).
\begin{figure}
    \includegraphics[width=0.485\textwidth]{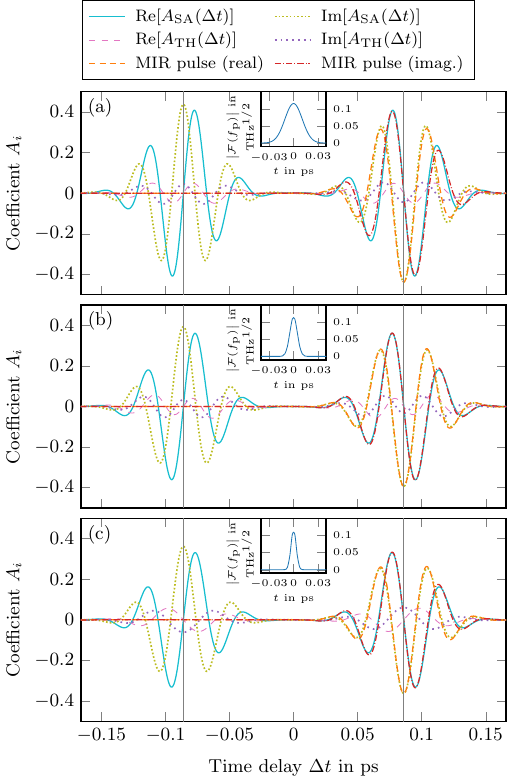}
	\caption{Time-delay dependence of the coefficients $A_{\text{SA}}(\Delta t)$ and $A_{\text{TH}}(\Delta t)$ describing the decomposition of the MIR mode function into sampled and thermalized contributions. The filtered probe is assumed to be of bandwidth $\Delta \tilde{\omega}/(2\pi) = \SI{1}{\tera\hertz}$ and central frequency $\tilde{\omega}/(2\pi) = \SI{300}{\tera\hertz}$, the nonlinear crystal is of length $L = \SI{100}{\micro\meter}$ and the refractive index can be taken from Eq.~\eqref{eq:refractive_index}. The sampled MIR pulse is a Gaussian of bandwidth $\sigma_{\tilde{\Omega}}/(2\pi) = \SI{5}{\tera\hertz}$ and central frequency $\tilde{\Omega}/(2\pi) = \SI{25}{\tera\hertz}$ and the probe of central frequency $\omega_{\text{p}}/(2\pi) = \SI{350}{\tera\hertz}$, amplitude $\alpha_\text{p} = 2\cdot 10^6$ and carrier envelope phase $t_{\text{p}} = 0$. The probe bandwidth is $\sigma_{\text{p}}/(2\pi) = \text{\SIlist{15;35;50}{\tera\hertz}}$ in (a)-(c). The dashed orange line shows the real and the dash-dotted red line the imaginary part of the MIR pulse shape in the time domain scaled and shifted to fit $A_{\text{SA}}(\Delta t)$. The vertical line marks the point $\Delta t = \pm\eta_{\text{c}}$.}
	\label{fig:coefficients}
\end{figure}
We can quantify how close the reconstructed variance and central frequency are to the corresponding MIR-pulse parameters by calculating the relative error for the variance
\begin{equation}\label{eq:rel_err_var}
	\frac{\abs{\sigma_{\tilde{\Omega}}^2 - \bar{\sigma}^2}}{\sigma_{\tilde{\Omega}}^2} = \frac{\sigma_{\tilde{\Omega}}^2}{\sigma_{\tilde{\Omega}}^2 + \sigma_{\text{p}}^2}
\end{equation}
and for the central frequency
\begin{equation}\label{eq:rel_err_freq}
	\frac{\abs{\tilde{\Omega} - \bar{\Omega}_{(-)}}}{\tilde{\Omega}} = \abs{1 + \frac{\tilde{\omega} - \omega_\text{p}}{\tilde{\Omega}}} \frac{\sigma_{\tilde{\Omega}}^2}{\sigma_{\tilde{\Omega}}^2 + \sigma_{\text{p}}^2}
.\end{equation}
We see that both relative errors vanish if the bandwidth of the pump pulse considerably exceeds the MIR pulse bandwidth, $\sigma_\text{p} \gg \sigma_{\tilde{\Omega}}$. We can also see that the relative error in the central frequency is always zero when the sampled frequencies are matched to the MIR, $\omega_\text{p} - \tilde{\omega} = \tilde{\Omega}$. Using Eq.~\eqref{eq:rel_err_var} and \eqref{eq:rel_err_freq} we can calculate the variance $\sigma_\text{p}$ of the pump pulse required to achieve a certain relative error in the reconstructed variance and central frequency given a fixed variance and central frequency of the MIR pulse. Some exemplary values can be taken from Table~\ref{tab:errors}.
\begin{table}[b]
	\centering
	\caption{Relative errors in the sampled central frequency and variance of the MIR waveform as well as the maximum of the coefficients $A_{\text{D/Th}}(\Delta t)$ in Fig.~\ref{fig:coefficients} for various values of the NIR pump pulse variance $\sigma_{\text{p}}$ and a fixed MIR pulse central frequency $\tilde\Omega = \SI{25}{\tera\hertz}$, variance $\sigma_{\tilde{\Omega}}/(2\pi) = \SI{5}{\tera\hertz}$, frequency mismatch $\omega_{\text{p}} - \tilde{\omega} = 2\tilde{\Omega}$ and a fixed carrier-envelope phase of the pump pulse $t_{\text{p}} = 0$.}
	\label{tab:errors}
	\begin{tabular}{c|c|c|c|c}
		 & relative error & relative error & maximum & maximum \\
		 $\sigma_\text{p}/(2\pi)$ &  variance & cent. freq. & $A_\text{SA}(\Delta t)$ & $A_\text{TH}(\Delta t)$ \\
		\hline
		\hline
		\SI{15}{\tera\hertz} & 10 \% & 10 \% & 0.44 & 0.02 \\
		\SI{35}{\tera\hertz} & 2 \% & 2 \% & 0.39 & 0.05 \\
		\SI{50}{\tera\hertz} & 1 \% & 1 \% & 0.36 & 0.06 
	\end{tabular}
\end{table}

\item \emph{Thermalization.}---As mentioned previously, there is an additional contribution to the signal due to the entanglement of the sampled mode with the thermalized mode $\hat{\mathfrak{a}}_{\text{TH}}$, which originates from the decomposition in Eq.~\eqref{eq:mir_mode_op_decomposition} and can be observed from Fig.~\ref{fig:coefficients}, where $A_{\text{TH}}(\Delta t)$ increases with the bandwidth of the pump pulse. One could try to reduce the contribution of the thermalized mode to the MIR mode, i.e., minimize the coefficient $A_{\text{TH}}(\Delta t)$ for all time delays $\Delta t$. A minimization of the coefficient $A_{\text{TH}}(\Delta t)$ in a mathematical sense seems not feasible since the coefficient depends on the waveform of the MIR mode. However, there are two strategies to mitigate the influence of the thermalized mode. First, Eq.~\eqref{eq:A_Th_approx} shows that for large pump bandwidths $\sigma_\text{p}$, $a_{(+)}(\Delta t)$ and $a_{(-)}(\Delta t)$ in Eq.~\eqref{eq:measured_pulse} oscillate coherently and for $\Phi_\perp = n\pi$ with $n \in \mathbb{Z}$ they interfere destructively. The latter requirement can be met by fixing the pump carrier-envelope to $t_\text{p} = 0$. Second, one can reduce the entanglement between the sampled and thermalized modes by decreasing the two-mode squeezing parameter $\zeta_{\text{T}}$. This can be achieved by filtering below the pump central frequency, $\tilde{\omega} - \omega_{\text{p}} \ll 0$, as can be seen from Fig.~\ref{fig:squeezing_params}. This way SFG detection is reduced, as can be seen in Fig.~\ref{fig:MIR_spektrum}, since less of the pump bandwidth is below the filtered frequency $\tilde{\omega}$. Coincidentally, filtering below the pump central frequency will reduce the single-mode squeezing and increase the overall squeezing parameter of the nonlinear interaction $\theta_{\tilde{\omega}}^{(1)}$, as can be observed in Fig.~\ref{fig:squeezing_params}. A mismatch between the central frequency of the sampled mode $\omega_\text{p} - \tilde{\omega}$ and the central frequency of the MIR mode $\tilde{\Omega}$ leads to a reduced up-conversion efficiency and in turn to a reduction of $A_{\text{SA}}(\Delta t)$, but since the pump pulse is rather broadband the reduction is minor. Combining the two strategies by mismatching the sampled frequencies of the setup and the MIR central frequency to $\omega_{\text{p}} - \tilde{\omega} = 2\tilde{\Omega}$ and fixing the carrier-envelope phase of the pump pulse, $t_{\text{p}} = 0$, the pump can be made short enough to allow for a subcycle resolution while keeping the influence of the thermalized mode down, as can be taken from Tab.~\ref{tab:errors}. For example, under such conditions a pump pulse with bandwidth $\sigma_\text{p}/(2\pi) = \SI{35}{\tera\hertz}$ can reach a relative error in the sampled central frequency and variance of $2\%$, while keeping the maximum of the $A_{\text{TH}}(\Delta t)$ below $13\%$ off from the maximum of $A_{\text{SA}}(\Delta t)$ [c.f. Fig.~\ref{fig:coefficients} (b)].

\item \emph{Deamplification.}---The overlap of the gating function and the MIR waveform is imperfect. From Fig.~\ref{fig:coefficients}, we can see that the maximum value of $A_{\text{SA}}(\Delta t)$ always remains well below one. The full reconstruction of the MIR quantum state requires the coefficient $A_\text{SA}(\Delta t)$ to be equal to one for at least one time delay $\Delta t$. Consider a coherent MIR sample state with amplitude $\alpha$ which would, if we assume no thermalization, appear to our measurement setup as a coherent state with amplitude $A_\text{SA}(\Delta t)\alpha$. Thus, to reconstruct $\alpha$, $A_\text{SA}(\Delta t)$ needs to be one for some $\Delta t$. In other words, the sampled mode function, defined by $f_{\text{SA}}(\Omega)$, has to coincide with that of the MIR pulsed mode, defined by $f_{\tilde{\Omega}}(\Omega)$. In this case, in the absence of any time delay, $\Delta t = 0$, the coefficient becomes one, $A_\text{SA}(\Delta t = 0) = \mathcal{F}[f_{\tilde{\Omega}}\conj{f}_{\text{SA}}](\Delta t = 0) = \int_0^\infty \abs{f_{\tilde{\Omega}}(\Omega)}^2 \dd \Omega = 1$. This can be achieved approximately by choosing $\sigma_\text{p} = \sigma_{\tilde{\Omega}}$ and $\omega_\text{p} - \tilde{\omega} = \tilde{\Omega}$. Furthermore, the nonlinear crystal has to be sufficiently small to suppress the influence of the phase matching, as can be seen from the approximation (and $\eta_{\text{c}} \propto L$)
\begin{equation}\label{eq:coefficient_approx}
	\abs{A_\text{SA}(\Delta t = 0)} \approx \exp(-\eta_\text{c}^2\sigma_{\tilde{\Omega}}^2/2)\sinc(\eta_\text{c}\tilde{\Omega})
,\end{equation}
which is valid as long as the pump pulse and MIR sample pulse spectrum overlap and dispersion is low in the MIR range. For $\sigma_{\tilde{\Omega}}/(2\pi) = \SI{5}{\tera\hertz}$, $(\omega_{\text{p}} - \tilde{\omega})/(2\pi) = \tilde{\Omega}/(2\pi) = \SI{25}{\tera\hertz}$ and $\tilde{\omega}/(2\pi) = \SI{300}{\tera\hertz}$, the nonlinear crystal has to be $L = \SI{6}{\micro\meter}$ for $\abs{A_\text{SA}(\Delta t)}$ to reach a maximum of 0.8 if the approximated expression in Eq.~\eqref{eq:coefficient_approx} is used (precise numerical evaluation of Eq.~\eqref{eq:A_D_approx} results in a maximum of 0.99). Numerical data on the coefficient $A_{\text{SA}}(\Delta t)$ for the previously mentioned values is shown in Fig.~\ref{fig:coeff_max}.
\end{enumerate}

\begin{figure}
    \includegraphics[width=0.485\textwidth]{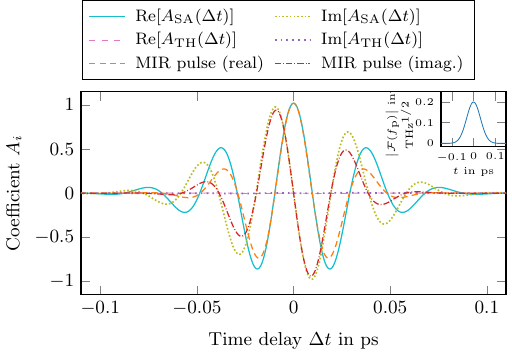}
	\caption{Time-delay dependence of the coefficients $A_{\text{SA}}(\Delta t)$ and $A_{\text{TH}}(\Delta t)$ describing the decomposition of the mode function into sampled and thermalized contributions for different time delays $\Delta t$. The filtered probe is of bandwidth $\Delta \tilde{\omega}/(2\pi) = \SI{1}{\tera\hertz}$, the central frequency is $\tilde{\omega}/(2\pi) = \SI{300}{\tera\hertz}$, and the refractive index given in Eq.~\eqref{eq:refractive_index}. In contrast to Fig.~\ref{fig:coefficients}, we consider a short nonlinear crystal with $L = \SI{6}{\micro\meter}$. Choosing a short crystal reduces the influence of the phase matching and allows a perfect match of the gating function $\mathcal{F}[\conj{f}_{\text{SA}}](\Delta t)$ and the MIR waveform $\mathcal{F}[f_{\tilde{\Omega}}](\Delta t)$. The MIR sample pulse is of bandwidth $\sigma_{\tilde{\Omega}}/(2\pi) = \SI{5}{\tera\hertz}$, central frequency $\tilde{\Omega}/(2\pi) = \SI{25}{\tera\hertz}$ and the probe of central frequency $\omega_{\text{p}}/(2\pi) = \SI{325}{\tera\hertz}$, amplitude $\alpha_{\text{p}} = 2\cdot 10^6$ and carrier envelope phase $t_{\text{p}} = 0$. The probe bandwidth is $\sigma_{\text{p}}/(2\pi) = \SI{5}{\tera\hertz}$.}
	\label{fig:coeff_max}
\end{figure}

\section{Conclusions}
In this work we propose an electro-optic-based tomography scheme able to reconstruct a time-dependent quantum state in the mid-infrared range, allowing for a dynamical sampling of spatiotemporally localized optical modes not only in phase space, but also in the time domain with subcycle resolution. Utilizing two recently developed theoretical tools \cite{Onoe2022,Hubenschmid2022}, we derive the photon-count probability distribution $p(\{\Delta n_i\})$ for multichannel subcycle quantum electro-optic sampling and show how this quantity is related to a specific time-dependent phase-space distribution of the sampled state. We scrutinize the physical sources of noise in the quantum electro-optic signal variance and show that sampling nonmonochromatic optical modes with subcycle resolution leads to two intriguing effects that detrimentally enhance the noise: A state-independent contribution scaling with the interaction strength between the sampled MIR and the detected NIR modes and thermalization noise due to entanglement breaking between the sampled and unsampled modes. The latter effect is unparalleled by continuous-wave-driven multichannel electro-optic sampling, which is mostly limited by noise due to the simultaneous measurement of noncommuting observables. We then propose a scheme to minimize the thermalization noise by a mismatch between the central frequencies of the pump pulse used to drive the nonlinear interaction and of the detected NIR modes. Finally, we demonstrate the reconstruction of the waveform using the example of a Gaussian mid-infrared pulsed mode. Our proposed optical tomography scheme is able to dynamically sample a broadband quantum state using an ultrabroadband pump pulse, opening a new paradigm for time-domain quantum tomography with subcycle resolution.

\begin{acknowledgments}
	We acknowledge funding by the Deutsche Forschungsgemeinschaft (DFG) - Project No. 425217212 - SFB 1432.
	T.L.M.G. gratefully acknowledges the funding by the Baden-Württemberg Stiftung via the Elite Programme for Postdocs. T.L.M.G. also acknowledges support by the ERC Starting Grant QNets through Grant No. 804247. 
\end{acknowledgments}

\appendix
\begin{widetext}
\section{The two-mode characteristic function}\label{a:two_mode_char}
In this section we derive the relation used in Eq.~\eqref{eq:cont_qpd} between the two-mode and single-mode characteristic function. The eigenstates of the operators $\{\hat{\mathfrak{u}}_{\tilde{\omega}_i} \mid i = \text{X},\text{Y}\} \cup \{\hat{\mathfrak{a}}_{\text{SA}},\hat{\mathfrak{a}}_{\text{TH}}\}$ span a Hilbert space $\hilbert_\text{EOS} = \left(\bigotimes_{i = \text{X},\text{Y}}\hilbert_{\tilde{\omega}_i}\right) \otimes \hilbert_{\text{SA}} \otimes \hilbert_{\text{TH}}$ which is a factor of the uncountably infinite dimensional Hilbert space $\hilbert = \hilbert_{\text{EOS}} \otimes \hilbert_{\text{r}}$. By defining the reduced density operator $\hat{\rho}_{\text{S-T}} = \tr_{\text{r}}(\hat{\rho}_{\tilde{\Omega}})$ of the modes, the electro-optic signal inherits contributions from (unsampled) modes correlated to the sampled ones. The density operator $\hat{\rho}_\text{S-T}$ acts on the effective bipartite Hilbert space $\hilbert_{\text{SA}} \otimes \hilbert_{\text{TH}}$ and can hence be written in terms of the displacement operators $\hat{D}_{\text{SA}}(\beta_{\text{SA}})$ and $\hat{D}_{\text{TH}}(\beta_{\text{TH}})$ generated by $\hat{\mathfrak{a}}_{\text{SA}}$ and $\hat{\mathfrak{a}}_{\text{TH}}$ (see \cite{Glauber2007}, p.~265):
\begin{align}\label{eq:mir_state}
	\hat{\rho}_\text{S-T} = \frac{1}{\pi^2} \int \chi_{\text{SA},\text{TH}}(\beta_{\text{SA}},\beta_{\text{TH}})\hat{D}_{\text{SA}}^\dagger(\beta_{\text{SA}})\hat{D}_{\text{TH}}^\dagger(\beta_{\text{TH}}) \dd^2 \beta_{\text{SA}} \dd^2 \beta_{\text{TH}}
,\end{align}
with the symmetrically ordered, two-mode characteristic function 
\begin{equation}\label{eq:two_mode_characteristic}
	\chi_{\text{SA},\text{TH}}(\beta_{\text{SA}},\beta_{\text{TH}}) = \ev{\hat{D}_{\text{SA}}(\beta_{\text{SA}})\hat{D}_{\text{TH}}(\beta_{\text{TH}})}_{\hat{\rho}_\text{S-T}}
.\end{equation}
However, as we have already seen, the effective nonlinear unitary operator~\eqref{eq:NL_effective} only acts on the subspace $\left(\bigotimes_{i \in I}\hilbert_{\tilde{\omega}_i}\right) \otimes \hilbert_{\text{SA}}$. Thus, we have to calculate the reduced density operator on the Hilbert space $\hilbert_{\text{SA}}$,
\begin{align}\label{eq:transformed_density_operator}
	\hat{\tilde{\rho}}_{\text{SA}} =\tr_{\text{TH}}\left[\hat{S}_{\text{S-T}}^\dagger\hat{S}_{\text{SA}}^\dagger \hat{\rho}_\text{S-T}\hat{S}_{\text{SA}}\hat{S}_{\text{S-T}}\right] = \frac{1}{\pi}\int \chi_{\text{SA}}(\beta) \hat{D}_{\text{SA}}^\dagger(\beta) \dd^2 \beta
,\end{align}
with the symmetrically ordered characteristic function $\chi_{\text{SA}}(\beta) = \ev{\hat{D}_{\text{SA}}(\beta)}_{\hat{\tilde{\rho}}_{\text{SA}}}$. Inserting Eq.~\eqref{eq:mir_state} into Eq.~\eqref{eq:transformed_density_operator} leads to
\begin{equation}
	\hat{\tilde{\rho}}_{\text{SA}} = \frac{1}{\pi^2} \int \chi_{\text{SA},\text{TH}}(\beta_{\text{SA}},\beta_{\text{TH}}) \tr_{\text{TH}}\left[\hat{S}_{\text{TH}}^\dagger\hat{S}_{\text{SA}}^\dagger \hat{D}_{\text{SA}}^\dagger(\beta_{\text{SA}})\hat{D}_{\text{TH}}^\dagger(\beta_{\text{TH}}) \hat{S}_{\text{SA}}\hat{S}_{\text{TH}}\right] \dd^2 \beta_{\text{SA}} \dd^2 \beta_{\text{TH}}
,\end{equation}
Apply the single-mode squeezing operator leads to
\begin{equation}
	\hat{S}_{\text{SA}}^\dagger \hat{D}_{\text{SA}}(\beta_{\text{SA}})\hat{S}_{\text{SA}} = \hat{D}_{\text{SA}}(\tilde{\beta}_{\text{SA}})
,\end{equation}
with $\tilde{\beta}_{\text{SA}} = \mu_{\text{SA}}\beta_{\text{SA}} - \nu_{\text{SA}}\conj{\beta_{\text{SA}}}$. Thus, with the Jacobian determinant $\det[D\tilde{\beta}_{\text{SA}}] = \mu_{\text{SA}}^2 - \abs{\nu_{\text{SA}}}^2 = 1$, we can write the reduced density operator as follows
\begin{equation}
	\hat{\tilde{\rho}}_{\text{SA}} = \frac{1}{\pi^2} \int \chi_{\text{SA},\text{TH}}(\mu_{\text{SA}}\tilde{\beta}_{\text{SA}} - \nu_{\text{SA}}\conj{\tilde{\beta}}_{\text{SA}},\beta_{\text{TH}}) \tr_{\text{TH}}\left[\hat{S}_{\text{TH}}^\dagger\hat{D}_{\text{SA}}^\dagger(\tilde{\beta}_{\text{SA}})\hat{S}_{\text{TH}}\hat{S}_{\text{TH}}^\dagger\hat{D}_{\text{TH}}^\dagger(\beta_{\text{TH}})\hat{S}_{\text{TH}}\right] \dd^2 \tilde{\beta}_{\text{SA}} \dd^2 \beta_{\text{TH}}
.\end{equation}
If we now apply the two-mode squeezing operators on the displacement operators
\begin{align*}
	&\hat{S}_{\text{TH}} \hat{D}_{\text{TH}}(\beta_{\text{TH}}) \hat{S}_{\text{TH}}^\dagger = \exp{\beta_{\text{TH}}\left[\mu_{\text{TH}}\hat{\mathfrak{a}}_{\text{TH}}^\dagger - \nu_{\text{TH}}\hat{\mathfrak{a}}_{\text{SA}}\right] - \hc} \\
	&= \hat{D}_{\text{TH}}(\mu_{\text{TH}}\beta_{\text{TH}}) \otimes \hat{D}_{\text{SA}}(\nu_{\text{TH}}\conj{\beta_{\text{TH}}}) \numberthis \\
	&\hat{S}_{\text{TH}} \hat{D}_{\text{SA}}(\tilde{\beta}_{\text{SA}}) \hat{S}_{\text{TH}}^\dagger = \exp{\tilde{\beta}_{\text{SA}}\left[\mu_{\text{TH}}\hat{\mathfrak{a}}_{\text{SA}}^\dagger - \nu_{\text{TH}}\hat{\mathfrak{a}}_{\text{TH}}\right] - \hc} \\
	&= \hat{D}_{\text{TH}}(\nu_{\text{TH}}\conj{\tilde{\beta}_{\text{SA}}}) \otimes \hat{D}_{\text{SA}}(\mu_{\text{TH}}\tilde{\beta}_{\text{SA}}) \numberthis
.\end{align*}
Using the fact, that the trace over the displacement operator gives a delta distribution (see p.339, \cite{Glauber2007})
\begin{align*}
	&\tr_{\text{TH}}\left\{\hat{D}_{\text{TH}}\left[\mu_{\text{TH}}\beta_{\text{TH}}\right]\hat{D}_{\text{TH}}\left[\nu_{\text{TH}}\conj{\left(\tilde{\beta}_{\text{SA}}\right)}\right]\right\} = \frac{\pi}{(\mu_{\text{TH}})^2}\delta\left[\Re(\beta_{\text{TH}}) + \frac{\nu_{\text{TH}}}{\mu_{\text{TH}}}\Re(\tilde{\beta}_{\text{SA}})\right] \delta\left[\Im(\beta_{\text{TH}}) - \frac{\nu_{\text{TH}}}{\mu_{\text{TH}}}\Im(\tilde{\beta}_{\text{SA}})\right] \numberthis
.\end{align*}
The reduced and transformed density operator can thus be written as follows,
\begin{align*}
	\hat{\tilde{\rho}}_{\text{SA}} =& \frac{1}{\pi^2} \iint \chi\left(\mu_{\text{SA}}\tilde{\beta}_{\text{SA}}-\nu_{\text{SA}}\conj{\tilde{\beta}_{\text{SA}}},\beta_{\text{TH}}\right) \\
	\times& \tr_{\text{TH}}\left[\hat{D}_{\text{TH}}\left(-\mu_{\text{TH}}\beta_{\text{TH}}\right)\hat{D}_{\text{TH}}\left(-\nu_{\text{TH}}\conj{\tilde{\beta}_{\text{SA}}}\right)\right] \hat{D}_{\text{SA}}\left(-\nu_{\text{TH}}\conj{\beta_{\text{TH}}}\right) \hat{D}_{\text{SA}}\left(-\mu_{\text{TH}}\tilde{\beta}_{\text{SA}}\right) \dd^2 \beta_{\text{TH}} \dd^2 \tilde{\beta}_{\text{SA}} \\
	=& \frac{1}{\pi\mu_{\text{TH}}^2} \int \chi\left(\mu_{\text{SA}}\tilde{\beta}_{\text{SA}}-\nu_{\text{SA}}\conj{\tilde{\beta}_{\text{SA}}},-\frac{\nu_{\text{TH}}}{\mu_{\text{TH}}}\conj{\tilde{\beta}_{\text{SA}}}\right) \hat{D}_{\text{SA}}\left(-\nu_{\text{TH}}\conj{\beta_{\text{TH}}}\right) \hat{D}_{\text{SA}}\left(-\mu_{\text{TH}}\tilde{\beta}_{\text{SA}}\right) \dd^2 \tilde{\beta}_{\text{SA}} \\
	=& \frac{1}{\pi} \int \chi\left[\mu_{\text{TH}}\left(\mu_{\text{SA}}\tilde{\beta}_{\text{SA}}-\nu_{\text{SA}}\conj{\tilde{\beta}_{\text{SA}}}\right),-\nu_{\text{TH}}\conj{\tilde{\beta}_{\text{SA}}}\right] \hat{D}_{\text{SA}}\left(-\tilde{\beta}_{\text{SA}}\right) \dd^2 \tilde{\beta}_{\text{SA}} \\
	=& \frac{1}{\pi} \int \chi(\beta) \hat{D}_{\text{SA}}(-\beta) \dd^2 \beta \numberthis 
.\end{align*}
Comparing the last two lines, we can conclude that 
\begin{equation}
    \chi(\beta) = \chi\left[\mu_{\text{TH}}\left(\mu_{\text{SA}}\beta-\nu_{\text{SA}}\conj{\beta}\right),-\nu_{\text{TH}}\conj{\beta}\right]
.\end{equation}

By applying a beam splitter after the spectral filtering to achieve the simultaneous measurement, the time evolution in Eq.~\eqref{eq:total_time_evolution} has to be modified to
\begin{equation}
	\hat{U} = \hat{U}_{\text{WP}} \hat{D}_{\tilde{\omega}}(\vec{\beta})\hat{U}_{\text{B}}\hat{U}_\text{NL}
,\end{equation}
with the operator $\hat{U}_{\text{B}} = \exp[\iu \frac{\pi}{4}(\hat{u}_{\tilde{\omega}_{\text{X}}}^\dagger \hat{u}_{\tilde{\omega}_{\text{Y}}} + \hc)]$. Since the beam splitter has no effect on the vacuum, it can be inserted into Eq.~\eqref{eq:prob_dist_def} resulting in
\begin{equation}
    p(\{\Delta n_i\}) = \tr(\hat{P}_{\{\Delta n_i\}} \hat{U}\hat{U}_{\text{B}^\dagger}\hat{\rho}_{\tilde{\Omega}} \otimes \ket{0}_\text{NIR}\prescript{}{\text{NIR}}{\bra{0}} \hat{U}_{\text{B}}\hat{U}^\dagger)
.\end{equation}
Applying the beam splitter to the nonlinear operator, $\hat{U}_{\text{B}}\hat{U}_\text{NL}\hat{U}_{\text{B}}^\dagger$, will result in a similar decomposition as Eq.~\eqref{eq:spectral_splitting} with $\tilde{\alpha}_{\text{X}} = 1/\sqrt{2}$ and $\tilde{\alpha}_{\text{Y}} = \iu/\sqrt{2}$. The additional imaginary unit can be accounted for by tuning the wave plates according to the result from Ref.~\cite{Hubenschmid2022}.

\section{The quasiprobability distribution for various states}\label{a:quasi_prob}
In this section we present the transformed $(s_\text{X},s_\text{Y})$-quasiprobability distributions, for different states. The $(s_\text{X},s_\text{Y})$-quasiprobability distributions are a generalization of the $s$-quasiprobability distributions (see Ref.~\cite{Hubenschmid2022} for details), which can be recovered through $s = s_\text{X} = s_\text{Y}$. 

Starting with a coherent MIR input state $\hat{\rho}_{\tilde{\Omega}} = \hat{D}_{\tilde{\Omega}}(\alpha_{\tilde{\Omega}})\ket{0}\bra{0}\hat{D}_{\tilde{\Omega}}^\dagger(\alpha_{\tilde{\Omega}})$, generated by $\hat{D}_{\tilde{\Omega}}(\alpha_{\tilde{\Omega}}) = \exp(\alpha_{\tilde{\Omega}}\int_0^\infty \conj{f}_{\tilde{\Omega}}(\Omega) \hat{a}_\Omega^\dagger \dd \Omega - \hc)$, and using $A_1(\Delta t) = \mu_{\text{TH}}\mu_{\text{SA}}A_{\text{SA}}(\Delta t)$, $A_2 = \mu_{\text{TH}}\conj{\nu_{\text{SA}}}A_{\text{SA}}(\Delta t) + \nu_{\text{TH}}A_{\text{TH}}(\Delta t)$ as well as
\begin{align}
	\sigma_{\text{X}}^2 =& \frac{1}{2}(\mu_\text{TH}^2\abs{\mu_\text{SA} - \nu_\text{SA}}^2 + \nu_\text{TH}^2 - s_\text{X}) \\
	\sigma_{\text{Y}}^2 =& \frac{1}{2}(\mu_\text{TH}^2\abs{\mu_\text{SA} + \nu_\text{SA}}^2 + \nu_\text{TH}^2 - s_\text{Y}) \\
	\sigma_{\text{XY}}^2 =& \mu_\text{TH}^2\mu_\text{SA}\abs{\nu_\text{SA}}\sin(\Phi_\perp)
,\end{align}
the quasiprobability distribution of a coherent state can be expressed as
\begin{align*}\label{eq:gaussian_state_qpd}
	\tilde{\rho}_{\text{SA}}(z;s_\text{X},s_\text{Y}|\alpha_{\tilde{\Omega}}) =& \left[\sigma_{\text{X}}^2\sigma_{\text{Y}}^2 - \sigma_{\text{XY}}^4\right]^{-\frac{1}{2}} \\
	&\times \exp\Big[\{-\Re^2([A_1(\Delta t) + A_2(\Delta t)]\alpha_{\tilde{\Omega}} - z)\sigma_{\text{X}}^2 - \Im^2([A_1(\Delta t) - A_2(\Delta t)]\alpha_{\tilde{\Omega}} - z)\sigma_{\text{Y}}^2 \\
    &- 2\Re([A_1(\Delta t) + A_2(\Delta t)]\alpha_{\tilde{\Omega}} - z)\Im([A_1(\Delta t) - A_2(\Delta t)]\alpha_{\tilde{\Omega}} - z)\sigma_{\text{XY}}^2\}/\{\sigma_{\text{X}}^2\sigma_{\text{Y}}^2 - \sigma_{\text{XY}}^4\}\Big]\numberthis
.\end{align*}
Next, let us assume the MIR modes are in a cat state $\hat{\rho}_{\tilde{\Omega}} = \ket{\text{cat}}\bra{\text{cat}}$ with $\ket{\text{cat}} = N_\text{cat}[\hat{D}_{\tilde{\Omega}}(\alpha_{\tilde{\Omega}}) + \hat{D}_{\tilde{\Omega}}(-\alpha_{\tilde{\Omega}})]\ket{0}$ and $N_\text{cat} = (2 + 2\e^{-2\abs{\alpha_{\tilde\Omega}}^2})^{-1/2}$, the transformed quasiprobability distribution is

\begin{align*}\label{eq:cat_state_qpd}
	\tilde{\rho}_{\text{SA}}(z;s_\text{X},s_\text{Y}|\text{cat}) =& 
    N_\text{cat}^2\Bigg(\tilde{\rho}_{\text{SA}}(z;s_\text{X},s_\text{Y}|\alpha_{\tilde{\Omega}}) + \tilde{\rho}_{\text{SA}}(z;s_\text{X},s_\text{Y}|-\alpha_{\tilde{\Omega}}) + 2\tilde{\rho}_{\text{SA}}(z;s_\text{X},s_\text{Y}|\text{vac})\exp\Big[-2\abs{\alpha_{\tilde{\Omega}}}^2 \\
    &+ \Big(\Re^2\big\{[A_1(\Delta t) - A_2(\Delta t)]\alpha_{\tilde{\Omega}}\big\}\sigma_\text{Y}^2 + \Im^2\big\{[A_1(\Delta t) + A_2(\Delta t)]\alpha_{\tilde{\Omega}}\big\}\sigma_\text{X}^2 \\
    &- 2\Re\big\{[A_1(\Delta t) - A_2(\Delta t)]\alpha_{\tilde{\Omega}}\big\}\Im\big\{[A_1(\Delta t) + A_2(\Delta t)]\alpha_{\tilde{\Omega}}\big\}\sigma_\text{XY}^2\Big)/(\sigma_{\text{X}}^2\sigma_{\text{Y}}^2 - \sigma_{\text{XY}}^4)\Big] \\
    &\times \cos\Big\{2\Big[
        \Big(\Im\big\{[A_1(\Delta t) + A_2(\Delta t)]\alpha_{\tilde{\Omega}}\big\}\sigma_{X}^2 - \Re([A_1(\Delta t) - A_2(\Delta t)]\alpha_{\tilde{\Omega}})\sigma_{\text{XY}}^2\Big)\Re(z) + \\
        & + (\Re([A_1(\Delta t) - A_2(\Delta t)]\alpha_{\tilde{\Omega}})\sigma_{Y}^2 - \Im([A_1(\Delta t) + A_2(\Delta t)]\alpha_{\tilde{\Omega}})\sigma_{\text{XY}}^2)\Im(z)
     \Big]/[\sigma_{\text{X}}^2\sigma_{\text{Y}}^2 - \sigma_{\text{XY}}^4]\Big\}\Bigg)\numberthis
.\end{align*}

If the MIR-input state is in the squeezed vacuum $\hat{\rho}_{\tilde{\Omega}} = \hat{S}_{\tilde{\Omega}}(\zeta_{\tilde{\Omega}})\ket{0}\bra{0}\hat{S}_{\tilde{\Omega}}^{\dagger}(\zeta_{\tilde{\Omega}})$, generated by
\begin{equation}\label{eq:mir_squeezing_op}
    \hat{S}_{\tilde{\Omega}}(\alpha_{\tilde{\Omega}}) = \exp[\frac{1}{2}\conj{\zeta}_{\tilde{\Omega}}\left(\int_{0}^{\infty} f_{\tilde{\Omega}}(\Omega) \hat{a}_{\Omega} \dd \Omega\right)^2 - \hc]
,\end{equation}
and using $A_1 = \mu_{\text{TH}}\mu_{\text{SA}}A_{\text{SA}}$, $A_2 = \mu_{\text{TH}}\nu_{\text{SA}}A_{\text{SA}} + \nu_{\text{TH}}A_{\text{TH}}$, as well as
\begin{align*}
	\sigma_{\text{X}}^2 =& \frac{1}{2}(\mu_{\text{TH}}^2\abs{\mu_{\text{SA}} + \nu_{\text{SA}}}^2 + \nu_{\text{TH}}^2 - s_\text{X}) + \Re(\mu_{\tilde{\Omega}}\conj{\nu_{\tilde{\Omega}}} (A_{1} - A_{2})^2) + \abs{\nu_{\tilde{\Omega}}}^2 \abs{A_{1} - A_{2}}^2 \numberthis \\
	\sigma_{\text{Y}}^2 =& \frac{1}{2}(\mu_{\text{TH}}^2\abs{\mu_{\text{SA}} - \nu_{\text{SA}}}^2 + \nu_{\text{TH}}^2 - s_\text{Y}) - \Re(\mu_{\tilde{\Omega}}\conj{\nu_{\tilde{\Omega}}} (A_{1} + A_{2})^2) + \abs{\nu_{\tilde{\Omega}}}^2 \abs{A_{1} + A_{2}}^2 \numberthis\\
	\sigma_{\text{XY}}^2 =& \mu_{\text{TH}}^2\mu_{\text{SA}}\abs{\nu_{\text{SA}}}\sin(\Phi_{\perp}) + \mu_{\tilde{\Omega}}[\Im(\conj{\nu_{\tilde{\Omega}}}A_{1}^2) - \Im(\conj{\nu_{\tilde{\Omega}}}A_{2}^2)] +2\abs{\nu_{\tilde{\Omega}}}^2\Im(A_{1}\conj{A_{2}}) \numberthis
,\end{align*}
the quasiprobability distribution takes a similar form to the coherent state, as the coherent states are squeezed by the interaction in the nonlinear crystal
\begin{align*}\label{eq:squeezed_vac_qpd}
	\tilde{\rho}_{\text{SA}}(z;s_\text{X},s_\text{Y}|\zeta_{\tilde{\Omega}}) =& \left[\sigma_{\text{X}}^2\sigma_{\text{Y}}^2 - \sigma_{\text{XY}}^4\right]^{-\frac{1}{2}} \\
	&\times \exp\Big[\{-\Re^2(z)\sigma_{\text{X}}^2 - \Im^2(z)\sigma_{\text{Y}}^2 - 2\Re(z)\Im(z)\sigma_{\text{XY}}^2\}/\{\sigma_{\text{X}}^2\sigma_{\text{Y}}^2 - \sigma_{\text{XY}}^4\}\Big]\numberthis
.\end{align*}

\section{Approximate solution to the coefficients}\label{a:coefficients}
In this section, we derive an approximate solution for the coefficients $A_{\text{SA}}(\Delta t)$ and $A_{\text{TH}}(\Delta t)$. The approximation is based on the assumption, that for the sampled MIR-frequency range, the refractive index of the nonlinear crystal is sufficiently flat. We use a simplified model for the refractive index of zinc telluride given by

\begin{equation}\label{eq:refractive_index}
    n_{\omega} = \Theta[b - \abs{\omega}/(2\pi)](a_1\abs{\omega} + c_1) + \Theta[\abs{\omega}/(2\pi) - b](a_2[\abs{\omega}-b]^2 + c_2)
\end{equation}
with $b=\SI{140}{\tera\hertz}$, $a_1 = \SI{3.5}{\cdot 10^{-4}\pico\second}$, $a_2 = \SI{2.6}{\cdot 10^{-6}\pico\second\squared}$, $c_1 = 2.55$, $c_2 = 2.75$ and $\Theta$ being the Heaviside step function. The model is based on data from Ref.~\cite{Leitenstorfer1999} for the MIR range and on Ref.~\cite{Marple1964} for the NIR. In the MIR range the refractive index is flat, as can be seen in Fig.~\ref{fig:refrative_index}.
\begin{figure}
    \includegraphics[width=0.6\textwidth]{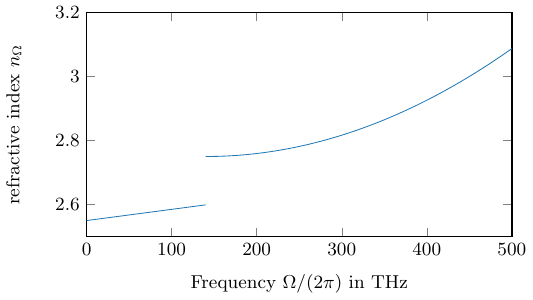}
	\caption{A simple model for the refractive index of zinc telluride. This was obtained by connecting the data from Ref.~\cite{Leitenstorfer1999} for the MIR range and data from Ref.~\cite{Marple1964} for the NIR. In the MIR range a linear relation was assumed, while in the NIR, it was assumed to be quadratic.}
	\label{fig:refrative_index}
\end{figure}
Using $\Omega n_{\Omega} = c k_{\Omega}$ we can expand $k_{\tilde{\omega} - \Omega} \approx k_{\tilde{\omega}} - \frac{\dd k_{\omega}}{\dd \omega} \vert_{\omega = \tilde{\omega}} \Omega = k_{\tilde{\omega}} - \frac{n_{\text{g}}(\tilde{\omega})}{c}\Omega$ to first order, where $n_{\text{g}}(\tilde{\omega})$ is the group refractive index at $\tilde{\omega}$, if $\abs{\frac{\dd k_\Omega}{\dd \Omega}}\vert_{\omega=\tilde{\omega}} \gg \frac{1}{2}\abs{\frac{\dd^2 k_\Omega}{\dd \Omega^2}\vert_{\omega=\tilde{\omega}}\Omega}$ for small $\Omega$ we can expand $(\tilde{\omega} - \Omega)n_{\tilde{\omega} - \Omega} = k_{\tilde{\omega} - \Omega} c$ to first order. Now we can approximate the frequency of the phase-matching function
\begin{equation}
	\eta_{\tilde{\omega},\Omega} \approx \frac{L\Omega}{2c}[n_{\text{g}} - n(\Omega)] \approx \frac{L\Omega}{2c}[n_{\text{g}}(\tilde{\omega}) - n(0)] = \eta_\text{c} \Omega
,\end{equation}
with the period of the phase-matching function $\eta_\text{c} = \frac{L}{2c}[n_\text{g}(\tilde{\omega}) - n(0)]$. Furthermore, using
\begin{equation}
	E_\text{p}(\omega) \approx \iu\left(\frac{\hbar}{4\pi c \varepsilon_0 A}\right)^{\frac{1}{2}}\sqrt{\frac{\abs{\omega_{\text{p}}}}{n_{\omega_{\text{p}}}}} f_\text{p}(\omega)
,\end{equation}
and defining
\begin{equation}\label{eq:S_c}
	S_\text{c} = -\frac{\iu}{\hbar} \left(\iu \sqrt{\frac{\hbar}{4\pi c \varepsilon_0 A}}\right)^3 \left(-4\pi L \frac{A \varepsilon_0 d}{2}\right) \sqrt{\frac{\tilde{\omega}}{n_{\tilde{\omega}}}} \conj{\alpha_\text{p}} \sqrt{\frac{\omega_\text{p}}{n_{\omega_\text{p}}}} \frac{1}{\eta_\text{c}}
,\end{equation}
the joint spectral amplitude can be written as
\begin{equation}\label{eq:S_approx}
	S(\Omega,\tilde{\omega}) \approx S_\text{c} f_\text{p}(\tilde{\omega} - \Omega)\sign(\Omega)\sqrt{\frac{\abs{\Omega}}{n(0)}}\frac{1}{\Omega}\sin(\eta_\text{c}\Omega) \numberthis
.\end{equation}
The above approximation of Eq.~\eqref{eq:S_approx}, allows us to analytically solve the integrals for the coefficients $A_\text{SA}(\Delta t)$ and $A_\text{TH}(\Delta t)$ defined through Eq.~\eqref{eq:mir_mode_op_decomposition} using $\erfcx(z) = \exp(z^2)[1-\erf(z)]$ resulting in
\begin{align*}
	A_{\text{SA}} \approx& \sech(\theta) \frac{\sqrt{\Delta \tilde{\omega}}}{\theta_{\tilde{\omega}}^{(1)}} S_\text{c}\e^{-\iu\tilde{\omega}t_\text{p}}\sqrt{\frac{1}{n(0)}} N_{\tilde{\Omega}} \frac{1}{2\iu} \sqrt{\pi} \bar{\sigma} \\
	&\times \exp[-(\omega_\text{p} - \tilde{\omega})^2/(4\sigma_\text{p}^2) - \tilde{\Omega}^2/(4\sigma_{\tilde{\Omega}}^2)] \Big\{\erfcx[\iu(\Delta t - \eta_\text{c})\bar{\sigma} - \bar{\Omega}_\text{SA}\bar{\sigma}] - \erfcx[\iu(\Delta t + \eta_\text{c})\bar{\sigma} - \bar{\Omega}_\text{SA}\bar{\sigma}]\Big\}, \numberthis \\
	A_{\text{TH}} \approx& \sec(\theta_{\perp})\csch(\theta) \frac{\sqrt{\Delta \tilde{\omega}}}{\theta_{\tilde{\omega}}^{(1)}} \conj{S}_\text{c}\e^{\iu\tilde{\omega}t_\text{p}}\sqrt{\frac{1}{n(0)}} N_{\tilde{\Omega}} \frac{1}{2\iu} \sqrt{\pi} \bar{\sigma} \\
	&\times \exp[-(\omega_\text{p} - \tilde{\omega})^2/(4\sigma_\text{p}^2) - \tilde{\Omega}^2/(4\sigma_{\tilde{\Omega}}^2)] \Big\{\erfcx[\iu(\Delta t - \eta_\text{c})\bar{\sigma} - \bar{\Omega}_\text{TH}\bar{\sigma}] - \erfcx[\iu(\Delta t + \eta_\text{c})\bar{\sigma} - \bar{\Omega}_\text{TH}\bar{\sigma}]\Big\} \\
	&- \tan(\theta_\perp)\e^{\iu\Phi_\perp} A_\text{SA} \numberthis
.\end{align*}
Since $f_{\tilde{\Omega}}(\Omega) \approx 0$ for $\Omega \leq 0$, we can expand the integral over the frequencies over the whole real numbers, resulting in the Fourier transform in Eq.~\eqref{eq:Convolution} and thus in Eq.~\eqref{eq:A_D_approx} and \eqref{eq:A_Th_approx}. Furthermore, we assume $\alpha_{\text{p}} \in \mathbb{R}$ and obtain $\Phi_\perp = \pi - 2\tilde{\omega}t_{\text{p}}$.

\section{Details to the thermalization noise}\label{a:therm}
Using the framework developed in Ref.~\cite{Hubenschmid2022}, we can relate the (ensemble) expectation values of the photon-count differences $\Delta n_i$ (for $i = \text{X},\text{Y}$) to the nonmonochromatic MIR quadratures $\hat{\mathfrak{X}}_{\text{SA}}(\varphi) = \frac{1}{2}(\hat{\mathfrak{a}}_{\text{SA}}\e^{\iu\varphi} + \hat{\mathfrak{a}}_{\text{SA}}^\dagger\e^{-\iu\varphi})$ with $\varphi_{\text{X}} = 0$ and $\varphi_{\text{Y}}= -\pi/2$ by
\begin{equation}\label{eq:expecation_val}
	\ev{\Delta n_{i}}_{\hat{\tilde{\rho}}_{\text{SA}}} \approx \sqrt{2}\abs{\nu}\abs{\beta_{Q}}\ev{\hat{\mathfrak{X}}_{\text{SA}}(\varphi_i)}_{\hat{\tilde{\rho}}_{\text{SA}}}
.\end{equation}
Similarly, the (root mean square) variances of the photon-number differences can be related to the variances of the quadratures, $\myvar{\hat{\mathfrak{X}}_{\text{SA}}(\varphi_i)} = \ev*{\hat{\mathfrak{X}}_{\text{SA}}^2(\varphi_i)}-\ev*{\hat{\mathfrak{X}}_{\text{SA}}(\varphi_i)}^2$, according to \cite{Hubenschmid2022}
\begin{equation}\label{eq:variance}
	\myvar{\Delta n_i} \approx 2\abs{\nu}^2\abs{\beta_i}^2\left(\myvar{\hat{\mathfrak{X}}_{\text{SA}}(\varphi_i)}\vert_{\hat{\tilde{\rho}}_\text{SA}} - \frac{\tilde{s}}{4}\right)
.\end{equation}
The variance has two contributions, one dependent on the MIR state and one only dependent on the parameter of the nonlinear interaction $\tilde{s} = 1 - 2\coth^2(|\theta_{\tilde{\omega}}^{(1)}|)$. The contribution from the latter can be reduced by increasing the squeezing parameter $\theta_{\tilde{\omega}}^{(1)}$, which can be achieved by increasing the amplitude $\alpha_\text{p}$ of the pump or by tuning the central frequencies $\tilde{\omega}$, $\omega_\text{p}$ according to Fig.~\ref{fig:squeezing_params}. The state-dependent contribution to the expectation value and the variance above are expressed in terms of the transformed sampled state of the MIR. However, they can be rewritten with respect to the state of the MIR mode $\hat{\rho}_{\tilde{\Omega}}$. Inserting Eq.~\eqref{eq:transformed_state} into the ensemble average and using the definitions $\hat{\vec{\mathfrak{X}}} = [\hat{\mathfrak{X}}_\text{SA}(\varphi),\hat{\mathfrak{X}}_\text{SA}(-\varphi - \Phi_\perp),\hat{\mathfrak{X}}_\text{TH}(-\varphi)]^\text{T}$ as well as $\vec{r} = [\mu_\text{T}\mu_\text{S},\mu_\text{T}\abs{\nu_\text{S}},\nu_\text{T}]^\text{T}$, the expectation value in Eq.~\eqref{eq:expecation_val} can be expressed as
\begin{equation}\label{eq:mir_quadrature}
	\ev{\hat{\mathfrak{X}}_{\text{SA}}(\varphi)}_{\hat{\tilde{\rho}}_{\text{SA}}} = \ev{\vec{r}^\text{T}\vec{\mathfrak{X}}}_{\hat{\rho}_{\tilde{\Omega}}}
,\end{equation}
and by defining the quantum covariance matrix 
\begin{equation}
	\cov(\hat{\vec{\mathfrak{X}}})\big\vert_{\hat{\rho}_{\tilde{\Omega}}} = \frac{1}{2}\ev{\left(\hat{\vec{\mathfrak{X}}} - \ev*{\hat{\vec{\mathfrak{X}}}}\right)\left(\hat{\vec{\mathfrak{X}}} - \ev*{\hat{\vec{\mathfrak{X}}}}\right)^\text{T} + \left[\left(\hat{\vec{\mathfrak{X}}} - \ev*{\hat{\vec{\mathfrak{X}}}}\right)\left(\hat{\vec{\mathfrak{X}}} - \ev*{\hat{\vec{\mathfrak{X}}}}\right)^\text{T}\right]^\text{T}}_{\hat{\rho}_{\tilde{\Omega}}}
,\end{equation}
with the following $(i,j)$-th matrix element $(\cov[\hat{\vec{\mathfrak{X}}}]_{\hat{\rho}_{\tilde{\Omega}}})_{ij} = \frac{1}{2}\ev*{\{\hat{\mathfrak{X}}_i,\hat{\mathfrak{X}}_j\}}_{\hat{\rho}_{\tilde{\Omega}}} - \ev*{\hat{\mathfrak{X}}_i}_{\hat{\rho}_{\tilde{\Omega}}}\ev*{\hat{\mathfrak{X}}_j}_{\hat{\rho}_{\tilde{\Omega}}}$ and the $i$-th/$j$-th component of $\hat{\vec{\mathfrak{X}}}$ the variance can be brought into the form
\begin{equation}\label{eq:var}
	\myvar{\hat{\mathfrak{X}}_{\text{SA}}(\varphi)}\big\vert_{\hat{\tilde{\rho}}_{\text{SA}}} = \vec{r}^\text{T}\cov(\hat{\vec{\mathfrak{X}}})\big\vert_{\hat{\rho}_{\tilde{\Omega}}}\vec{r} = \cov(\vec{r}^\text{T}\hat{\vec{\mathfrak{X}}})\big\vert_{\hat{\rho}_{\tilde{\Omega}}}
.\end{equation}
The first summand of the three contributions to the observable $\vec{r}^\text{T}\hat{\vec{\mathfrak{X}}}$ can be understood as follows.
Since the refractive index is flat in the MIR-range, the phase-matching function in Eq.~\eqref{eq:phase_matching} can be simplified to $\sinc(\eta_{\text{c}}\Omega)$ with $\eta_\text{c} = \frac{L}{2c}[n_\text{g}(\tilde{\omega}) - n(0)]$ depending on the group refractive index $n_{\text{g}}(\tilde{\omega})$ (see Appendix~\ref{a:coefficients} for details). The spectral function of the sampled mode $f_{\text{SA}}(\Omega)$ is therefore proportional to the pump spectral function modulated by the phase-matching function $\sinc(\eta_{\text{c}}\Omega)$. If $\abs{f_{\text{SA}}(\Omega)}$ (and thus the pump bandwidth, $\sigma_{\text{p}}$) is much broader than $\abs{f_{\tilde{\Omega}}(\Omega)}$, the expectation value can be approximated by 
\begin{align*}
    \tr(\hat{\mathfrak{\mathfrak{X}}}_{\text{SA}}(\varphi)\hat{\rho}_{\tilde{\Omega}}) \underset{\sim}{\propto}& 
    \tr(\hat{X}_{t_{\text{p}} - \eta_{\text{c}}}(\tilde{\varphi})\hat{\rho}_{\tilde{\Omega}}) + \tr(\hat{X}_{t_{\text{p}} + \eta_{\text{c}}}(\tilde{\varphi})\hat{\rho}_{\tilde{\Omega}}) \numberthis
\end{align*}
with $\tilde{\varphi} = \varphi + \arg[f_{\text{SA}}(\tilde{\Omega})]$ at the central frequency $\tilde{\Omega}$ of the MIR and the instantaneous quadrature $\hat{X}_{t}(\tilde{\varphi})$ at time $t$. In the limit case of an ideal classical electro-optic measurement with infinitely short pump pulses, the signal is directly related to the instantaneous quadrature expectation values, since the MIR pulse is sampled over a time slice defined by the pump pulse duration. A small uncertainty in time $\sigma_t$ will lead to a broader frequency band $\sigma_\omega$ of the probing system (in the present case, the pump $E_\text{p}$) due to the Gabor limit $\sigma_\omega^2 \geq \pi/(2\sigma_t^2)$ \cite{Gabor1946}. However, as can be seen from Fig.~\ref{fig:MIR_spektrum}, a broader banded pump pulse will lead to some SFG in addition to the sampled DFG contribution since the higher end of the band is above the filtered frequency $\tilde{\omega}$ but some of the lower end of the band is below $\tilde{\omega}$. This will lead to an increase of the single-mode and two-mode squeezing parameter $\zeta_\text{S}$, $\zeta_\text{T}$ accounting for the SFG contributions, as can be seen in Fig.~\ref{fig:squeezing_params}. The two-mode squeezing creates entanglement between the sampled MIR mode and a temporal mode, which is not sampled and therefore leads to entanglement breaking which in turn mixes the sampled state and thus increases its von Neumann entropy, which is referred to as thermalization. As will be shown in the following, for a coherent MIR input state $\hat{\rho}_{\tilde{\Omega}}$ (including the vacuum as a limit case), the main result from Ref.~\cite{Onoe2022} can be reproduced using Eq.~\eqref{eq:var}, which shows that in the squeezing regime, i.e., $\overline{\mathfrak{a}}_{\tilde{\omega}} = \hat{\mathfrak{a}}_{\tilde{\omega}}^\dagger$, thermalization leads to an increase of the variance. Expanding Eq.~\eqref{eq:var} leads to
\begin{align*}
	\myvar{\hat{\mathfrak{X}}_{\text{SA}}(\varphi)}\big\vert_{\hat{\tilde{\rho}}_{\text{SA}}} =& \vec{r}^\text{T}\cov(\hat{\vec{\mathfrak{X}}})\big\vert_{\hat{\rho}_{\tilde{\Omega}}}\vec{r} \\
	=& \Bigg\{\mu_\text{TH}^2\mu_\text{SA}^2\sigma_{\hat{\mathfrak{X}}_\text{SA}(\varphi)}^2 + \mu_\text{TH}^2\abs{\nu_\text{SA}}^2\sigma_{\hat{\mathfrak{X}}_\text{SA}(-\varphi - \Phi_\perp)}^2 + \nu_\text{TH}^2\sigma_{\hat{\mathfrak{X}}_\text{TH}(-\varphi)}^2 + 2\mu_\text{TH}^2\mu_\text{SA}\abs{\nu_\text{SA}}\cov[\hat{\mathfrak{X}}_\text{SA}(\varphi),\hat{\mathfrak{X}}_\text{SA}(-\varphi-\Phi_\perp)] \\
	&+ 2\mu_\text{TH}\nu_\text{TH}\abs{\nu_\text{SA}}\cov[\hat{\mathfrak{X}}_\text{SA}(-\varphi-\Phi_\perp),\hat{\mathfrak{X}}_\text{TH}(-\varphi)] + 2\mu_\text{TH}\nu_\text{TH}\mu_\text{SA}\cov[\hat{\mathfrak{X}}_\text{SA}(\varphi),\hat{\mathfrak{X}}_\text{TH}(-\varphi)]\Bigg\} \Bigg\vert_{\hat{\rho}_{\tilde{\Omega}}} \numberthis
.\end{align*}
If we assume, that the reduced density operator $\hat{\rho}_{\tilde{\Omega}}(\hat{a}_{\tilde{\Omega}}) = \hat{\rho}_{\text{SA}}(A_\text{SA}\hat{a}_{\text{SA}})\otimes\hat{\rho}_{\text{TH}}(A_\text{TH}\hat{a}_{\text{TH}})\otimes\hat{\rho}_{\text{U}}(A_\text{U}\hat{a}_{\text{U}})$ separates, 
\begin{align*}
	\myvar{\hat{\mathfrak{X}}_{\text{SA}}(\varphi)}\big\vert_{\hat{\tilde{\rho}}_{\text{SA}}} =& \Bigg\{\mu_{\text{TH}}^2\Big[\mu_{\text{SA}}^2\myvar{\hat{\mathfrak{X}}_{\text{SA}}(\varphi)} + |\nu_{\text{SA}}|^2\myvar{\hat{\mathfrak{X}}_{\text{SA}}(-\varphi)}\Big] + \nu_{\text{TH}}^2\myvar{\hat{\mathfrak{X}}_{\text{TH}}(-\varphi)} \\
	&+ \mu_{\text{TH}}^2\mu_{\text{SA}}|\nu_{\text{SA}}|\Big(\ev{\{\hat{\mathfrak{X}}_{\text{SA}}(\varphi),\hat{\mathfrak{X}}_{\text{SA}}(-\varphi-\Phi_{\perp})\}} - 2\ev{\hat{\mathfrak{X}}_{\text{SA}}(\varphi)}\ev{\hat{\mathfrak{X}}_{\text{SA}}(-\varphi-\Phi_{\perp})}\Big)\Bigg\}_{\hat{\rho}_{\tilde{\Omega}}} \numberthis
.\end{align*}
For a coherent state $\hat{\rho}_{\tilde{\Omega}}(\hat{a}_{\tilde{\Omega}}) = \hat{D}_{\tilde{\Omega}}(\alpha_{\tilde{\Omega}})\ket{0}\bra{0}\hat{D}^\dagger_{\tilde{\Omega}}(\alpha_{\tilde{\Omega}})$, the variances are
\begin{align*}\label{eq:var_coh}
	\sigma_{\hat{\mathfrak{X}}_{\text{SA}}(\varphi)}^2\vert_{\hat{\tilde{\rho}}_{\text{SA}}} &= \mu_{\text{TH}}^2 \frac{1}{4} (\mu_{\text{SA}}^2 + \abs{\nu_{\text{SA}}}^2) + \frac{1}{4}\nu_{\text{TH}}^2 + \e^{2\iu \varphi}\mu_{\text{TH}}^2\mu_{\text{SA}}\abs{\nu_{\text{SA}}}\frac{1}{2}\cos(\Phi_\perp) \\
	&= \frac{1}{2}\ev{\hat{\mathfrak{a}}_{\tilde{\omega}}^\dagger \hat{\mathfrak{a}}_{\tilde{\omega}}} \pm \frac{1}{2}\Re(\ev{\hat{\mathfrak{a}}_{\tilde{\omega}}^2}) - \frac{1}{4} \numberthis
.\end{align*}
If $\hat{\mathfrak{X}}_{\text{SA}}(\varphi_\text{X})$ and $\hat{\mathfrak{X}}_{\text{SA}}(\varphi_\text{Y})$ are measured separately and thus $\tilde{s} = \sech^2(\theta_{\tilde{\omega}}^{(1)})$, the result from \cite{Onoe2022} is reproduced. As explained in the main text, it is still possible to mitigate the thermalization noise without compromising on the bandwidth of the pump, by filtering below pump central frequency, i.e., increasing the difference $\tilde{\omega} - \omega_{\text{p}}$, thus sampling less SFG since more of the bandwidth is above the filtered frequency $\tilde{\omega}$. As a result, the single-mode and two-mode squeezing is reduced, which agrees with Fig.~\ref{fig:squeezing_params}. In the next section, we will make use of this fact to enable the reconstruction of a quantum states waveform on a subcycle scale using the electro-optic sampling describe in the previous section.

\end{widetext}

\bibliography{lit_time_domain_QST}

\end{document}